# Equation of State based on the first principles

Sergey G. Chefranov

Physics Department, Technion-Israel Institute of Technology, 32000, Haifa, Israel

csergei@technion.ac.il

## Abstract

An alternative to the well-known complete form of the Mie-Grüneisen equation of state (EOS) for water is suggested. A closed analytical description of the self-consistent EOS for an arbitrary medium based only on the first law of thermodynamics and on a new form of virial theorem is obtained. This form of the virial theorem (allowing a variable power-law exponent of the particles interaction potential) is a result of the generalization of the known method of similarity (Feynman et al., 1949). In the new EOS, the description of the internal potential energy as a solution of a nonlinear Riemann-Hopf type equation is proposed.

## 1. Introduction

In order to solve fundamental and applied problems of high pressure and temperature physics, as well as problems of hydrodynamics, plasma physics and astrophysics, the equation of state (EOS) of the medium is used [1]-[17]. The EOS, which is usually a functional relationship of pressure with density and temperature [1], is used as an additional equation that makes it possible to compose a closed system of equations of hydrodynamics for a compressible medium [2]-[4]. EOS is also interesting by itself, giving a description of stationary, equilibrium and locally equilibrium states of matter [1], [5], [6]. In particular, the EOS applies to study of the extreme states of matter at ultrahigh pressures and temperatures characteristic of the inner regions of planets and other space objects, as well as for energy technical systems [16]. The use of the EOS gives certain advantages in comparison with numerical methods of molecular dynamics, naturally complementing them when analyzing the states of the medium taking into account the interaction of particles [4]-[6].

Due to the absence of a small parameter characterizing the strong interaction of particles of the medium at its extreme states, the problem of obtaining an analytical EOS based only on the first principles remains unsolved. The existing methods for obtaining the EOS using virial decompositions of pressure by degrees of density are based on the numerical determination of the temperature dependent fitting functions for the corresponding decomposition coefficients [3], [5], [6].

Similarly, in [2], [3] it is necessary to use additional data from experimental observations of shock wave parameters to reconstruct the EOS based on the integral form of the equations of hydrodynamics of a compressible medium in the form of Rankine-Hugoniot relations.

Besides, the additional use of experimental data is need in [15], [17], when an EOS for water in the complete Mie-Grüneisen form is considered. For example, a wide-range analytical EOS for water introduced in [17] can be applied at high pressures, which can be used to solve the problem of fusion ignition [16].

The original Mie-Grüneisen EOS [18]-[21] is sometimes said to be an incomplete one, due to the fact that it is expressed in such a way that the pressure becomes a function of a volumetric strain measure and of the internal energy with no possibility to define neither temperature nor entropy. In order to make it a complete EOS, the other form is to be specified, in which the pressure and the internal energy are a functions of the density and temperature.

Historically, the original Mie-Grüneisen EOS has been primarily used in assessing compression of metal solids. It assumes that the pressure $p$ is non-linearly related to the density $\rho$ but linear in the internal energy density (on the unit mass) $u$, when the representation of pressure is as follows $p - p_0(\rho) = \rho \Gamma(\rho)(u - u_0(\rho))$ [18]-[23]. It may be obtained after the integration of the Grüneisen relation $\Gamma(\rho) = \frac{1}{\rho} \left( \frac{\partial p}{\partial u} \right)_{\rho = const}$, in which the Grüneisen parameter $\Gamma$ is introduced as not dependent from pressure and energy density when sometimes relation $\Gamma(\rho)\rho = \Gamma_0 \rho_0 = const$ is also suggested [22]. In this incomplete form of the Mie-Grüneisen EOS the arbitrary functions $u_0$ and $p_0$ of such integration are determined at the compressive reference curve which can be a Hugoniot one [22].

When in addition to the Grüneisen relation the specific heat capacity is assumed to have the constant value $c_V = const$, the next complete form $p = \rho \Gamma c_V T + p_0(\rho)$, $u = c_V T + u_0(\rho)$ of the original Mie-Grüneisen EOS is stated. In this complete EOS, functions $u_0(\rho)$ and $p_0(\rho)$ are related through the first and the second laws of thermodynamics in the following form: $p_0(\rho) = \rho^2 du_0 / d\rho$ [20] (see also [7]).

The Mie-Grüneisen model, because of its simplicity and analyticity, is probably the most-used thermal EOS in high-pressure physics not only for solids, but also for any condensed matter. However, the usual method of implementing this model places unnecessary constraints on temperature calculations and fails to take advantage of information that can be obtained from the compression behavior of a material. As multiphase models requiring explicit accounting for temperature become more common, these issues suggest that a return to an internal energy in the more common representations as the starting point for an EOS model is useful [17], [21]-[23].

Thus, in the complete EOS of the Mie-Grüneisen type, the total pressure $p$ and the total internal energy (per unit mass) $u$ are described as the sum of thermal and mechanical components of the energy with linear representation of the thermal part of pressure through the thermal part of the internal energy [17]: $p = p_\rho(\rho) + p_T(\rho; T)$, $u = u_\rho(\rho) + u_T(\rho; T)$, where $\rho = mN / V; u = E / Nm$ and $N$ is the number of particles of a mass $m$ in a volume $V$. Here in the linear relation $p_T = \rho \Gamma_T u_T$, the Grüneisen coefficient $\Gamma_T(T; \rho)$ in common case is different from the Grüneisen parameter $\Gamma$ and determines the integral relation between the thermal part of the pressure and the thermal part of the internal energy $u_T$ [17].

In such a complete and wide-range form of the Mie-Grüneisen equation of state, it is already assumed that there is a possibility to take into account the temperature dependence for the Grüneisen coefficient $\Gamma_T(\rho; T)$, in contrast to the above root form of the Mie-Grüneisen equation of state. However, at the same time, it is necessary to inevitably use the temperature dependence of the specific heat capacity $c_V$ obtained from the observational data [15], [17]. This is a consequence of the unclosed system of equations for three unknown functions $\Gamma_T(T; \rho); p_T(T; \rho)$ and $u_T(T; \rho)$, since only two equations are used to determine them. One of them is a linear coupling equation $p_T = \rho \Gamma_T u_T$, and the second one follows from the first law of thermodynamics and has the well - known form (see Eq. (5) below and Eq. (16.5) in [1] or Eq. (15) in [17]).

As a result, well-known approaches to determining the EOS, including its tabular form [14], also leave open a number of problems of the hydrodynamics of a compressible medium, where the use of an analytical explicit form of the EOS is required [8]-[13]. Up to now it is difficult to separate the influence of the non-ideal behavior in the thermodynamic functions in the interpretation of the existing experimental data due to the various theoretical assumptions that are introduced besides first principles.

In our paper, a closed analytical EOS based on the use of the well-known virial theorem [1] is obtained. In particular, from the first principles the complete Mie-Grüneisen EOS is stated. Thus, there is no need in additional experimental data as usually to obtain the complete form of the Mie-Grüneisen EOS.

The significance of the virial theorem is that it allows the average kinetic and potential energy to be calculated for very complicated systems and holds even for systems that are not in thermal equilibrium [1], [25], [26]. The well-known van der Waals EOS was obtained as a first application of the virial theorem [26]. Virial theorem also was used in [27] to obtain an analytical form of the EOS for water and steam near the critical point. An analytical relationship between pressure, specific volume and temperature were received in [27], based on the virial theorem in the form of a statistical integral depending on the type of potential of intermolecular interaction (see also [4]-[6]). However, this approach does not offer an explicit analytical form of EOS in the form of Mie-Grüneisen to describe the parameters of strongly compressed water behind the shockwave front because in general case, the potential energy of the intermolecular interaction must be expressed in terms of macroscopically measured parameters of the medium.

Meanwhile, the latter can be done by using the virial theorem with the assumption of uniformity of the function, describing the potential of intermolecular interaction [1]. Only then can an explicit kind of connection between internal energy and pressure be established when using this uniformity exponent $d$, which is usually assumed to be constant and independent of density and temperature [1].

In this paper, due to the rejection of this hitherto generally accepted assumption about the constancy of the uniformity exponent $d$, a generalization of the theory [28] is proposed, which uses the similarity transformation method and the integral form of the first law of thermodynamics to derive the virial theorem. Thus, a restriction is obtained on the type of function $d(\rho;T)$, which follows from the compatibility condition of the above differential form of the first law of thermodynamics (7) with the new form of the virial theorem derived for an arbitrary type of function $d(\rho;T)$. This constraint has the form of a nonlinear Riemann-Hopf type equation (7), which admits a general exact explicit analytical solution for the function $d(\rho;T)$ (see Eq. (9) below).

In Section 2 a closed description for the internal potential energy and corresponding EOS is obtained in the form of an exact analytical solution of the nonlinear Riemann-Hopf type equation. In Section 3, an analytical form for the Grüneisen coefficient and internal energy, which allows to a self-consistent description of the EOS of water behind the strong shock wave front is obtained. In Section 4, the dependence of the isothermal EOS exponent for water on the Grüneisen parameter and on the power-law exponent of the interaction energy between water molecules for different compression is shown. In Section 5 the discussion about the EOS applications is represented. In Appendix A the new derivation of the virial theorem which gives the generalization of Feynman's et al. [28] approach and examples of exact solutions for EOS are represented. In Appendix B the closure problem in the compressible fluid dynamics is considered in the connection of the EOS using. In Appendix C the Grüneisen parameter derivation is obtained.

## 2. Analytical self-consistent EOS

In [28] to obtain the EOS based on the generalized Thomas-Fermi theory the virial theorem is derived by using similarity considerations and the first law of thermodynamics.

The consideration in [28] is limited only to the case of the Coulomb interaction potential proportional to $O(1/r)$, as required when formulating the boundary condition in the limit $r \to 0$ for the Thomas-Fermi model. However, when considering the pare interaction of charged particles in a dielectric, it is already possible to have exponents that do not coincide with the monopole Coulomb potential for the corresponding interaction potential. In this case, the potential energy of the interaction can already be proportional to $r^{-\tilde{d}}$ in the limit $r \to 0$, when the dependence of the exponent $\tilde{d} = \tilde{d}(r)$ on the distance to the point source of the field appears. This is analogous to the introduction of the dependence of the charge magnitude on the distance due to the polarization of the vacuum, considered as an analog of the dielectric medium [29]. At the same time, the case when $\tilde{d}(r) \to 0$ in the limit $r \to 0$, is also similar to the effect of asymptotic freedom [30].

In this regard, here in the first Subsection of Appendix A, a generalization of the consideration of the method of work [28] for the specified more general type of interaction potential is proposed and the conclusion of the virial theorem for the case $\tilde{d} \neq const$ is obtained.

Thus, one can express the average potential energy of the system in terms of its macroscopic parameters (pressure, density and temperature) by using the virial theorem and the first law of thermodynamics and so one can obtain self-consistent EOS for water or any other condensed matter.

For this purpose let us consider a finite non-periodic system of $N$ particles, each of mass $m$ that confined to a volume $V = mN / \rho$ when the internal potential energy $E_P(\vec{r}_1, \vec{r}_2, ..., \vec{r}_N) / mN = u_P$ is determined only by the interaction between particles. For the homogeneous function $u_P$ that admits relation $u_P(\lambda \vec{r}_1; ...; \lambda \vec{r}_N) = \lambda^{-d} u_P(\vec{r}_1; ...; \vec{r}_N)$ a well-known equation $\sum_{\alpha=1}^{N} \vec{r}_\alpha \frac{\partial u_P}{\partial \vec{r}_\alpha} = -d u_P$ is proposed on the base of the Euler theorem [1]. In case of a variable value of the uniformity exponent $d$ that equation is also valid for locally uniformity of functions $u_P$ with the uniformity exponent $d$ for which relation $u_P(\lambda \vec{r}_1; ...; \lambda \vec{r}_N) = \lambda^{-d} u_P(\vec{r}_1; ...; \vec{r}_N)$ is fulfilled only in the limit $\lambda - 1 \ll 1$ (see the second Subsection of Appendix A). This gives the generalization of Euler's theorem, which was considered earlier only in case of globally uniformity of functions with a constant value of the uniformity exponent $d = const$.

For example, in the simple case of pare interactions potential the variable value of the potential energy uniformity exponent is generally represented as (see (A5) in Appendix A) in the following form: $d = \tilde{d}(r) + r \ln r \frac{d}{dr} \tilde{d}(r)$. Thus, the virial theorem, at an arbitrary form of dependence $\tilde{d}$ on $r$, includes a value $d$, that may be also associated with the power-law exponent $\tilde{d}$ of the potential for pare interaction.

Let us consider the common representation for the potential energy which is analogous to the well-known form used [18], [20] for constant value of exponent $d$:

$$u_P / u_0 = (\rho / \rho_0)^{d(\rho;T)/3}, \qquad (1)$$

In (1) $u_0; \rho_0; r_0$ - arbitrary constants and the variable uniformity exponent $d$ of the potential energy $u_P$ is arbitrary function of density and temperature when all possible interactions between particles may be taken into account. Opposite to representation used in [18] and [20], in (1) the variable function $d$ on temperature and density is introduced. Meanwhile, in (1) the well-known relation between an average distance between the particles of the medium and its density $r / r_0 = (\rho / \rho_0)^{-1/3}$ (see also (17) in [20] and (3) in [31]) is used. For example, the specified form of potential energy with a positive value of a constant coefficient $u_0 > 0$ is intended for modeling the mutual repulsion of particles when they significantly approach in a compressed condensed medium, for example, behind the front of a strong shock wave. At the same time, we may omit specifying the physical mechanism of such repulsion, which allows us to expand the scope of applicability of the theory proposed below.

It is convenient for further consideration to represent from (1) the unknown function $d$ as a function of the potential energy and density of the medium in the following form:

$$d = \frac{3\ln(u_P / u_0)}{\ln(\rho / \rho_0)} \qquad (2)$$

At the same time, unlike the well-known virial method of decompositions [4]-[6] by integer degrees of density, the form of the function $d(\rho;T)$ is determined further without any additional assumptions.

Taking into account (2) and the well-known representation of the virial theorem [1], we obtain a representation for pressure (see also (A9) in Appendix A):

$$p = \frac{kT}{m}\rho + \rho d(\rho; u_P) u_P / 3 = \frac{kT}{m}\rho + \rho u_P \frac{\ln(u_P / u_0)}{\ln(\rho / \rho_0)} \qquad (3)$$

Let us use the definition of the energy density $u = E / mN = (E_K + E_P) / mN$ in the following form:

$$u = \frac{3kT}{2m} + u_P(\rho; T) \qquad (4)$$

An equation supplementing equation (3) may be obtained from the first law of thermodynamics in the well-known form [1], [17] (see also (1.10) in [22])

$$T\left(\frac{\partial p}{\partial T}\right)_\rho - p = -\rho^2 \left(\frac{\partial u}{\partial \rho}\right)_T \qquad (5)$$

After substituting representations (3), (4) in (5) and introducing new variables, we obtain a Riemann type equation [32], [33], which describes the dependence of the potential energy density on density and temperature:

$$\frac{\partial w}{\partial \tilde{y}} + (w+1)\frac{\partial w}{\partial x} - w = 0;$$

$$w = \ln(u_p / u_0) \equiv \frac{d(\tilde{y};x)}{3}\ln\frac{\rho}{\rho_0};$$

$$\tilde{y} = \ln\ln(\rho / \rho_0); x = \ln(T / T_0) \qquad (6)$$

In (6), for the introduction of a dimensionless unknown function $w$ and variables $x, \tilde{y}$, the corresponding arbitrary constant values are introduced, the values of which must be determined by taking into account the boundary or initial conditions.

In Appendix A an exact analytical solution in the implicit form is obtained by using of the Lagrange-Sharpie method in the common (see Eq. (A16) in the Subsection 3 of Appendix A). Also the more detail consideration of the simple but nontrivial solution to Eq. (6) in the limit $w \ll 1$ is also represented in Eq. (A19) and Eq. (A20) of Appendix A. In particular, based on the solution (A 19), representations are obtained for the second virial coefficient (A 21) and for the corresponding Boyle temperature (A22).

Let us consider an explicit form of the common solution for equation (6) in the opposite limit $w \gg 1$, when equation (6) has the following form:

$$\frac{\partial w}{\partial \tilde{y}} + w\frac{\partial w}{\partial x} - w = 0 \qquad (7)$$

Equation (7) exactly coincides with the modification of the Riemann-Hopf equation considered in [32], [33], where its exact solution is obtained.

Thus, a general exact explicit analytical solution of equation (7) has a known form [32], [33]:

$$w(x;y) = y\int_{-\infty}^{\infty} d\xi \overline{w}_0(\xi)\left(1 + y\frac{d\overline{w}_0}{d\xi}\right)\delta\left(\xi - x + y\overline{w}_0(\xi)\right); y = \exp \tilde{y} = \ln\frac{\rho}{\rho_0} \qquad (8)$$

In (8), the Dirac delta function is used under the integral, and the solution itself is obtained using the representation $w(x;y) = y\overline{w}(x;y); \overline{w}_0(x) \equiv w(x;y=0)$.

Based on solution (8), we obtain the following explicit exact solution for the exponent of the potential energy:

$$d(\rho;T) = \int_{-\infty}^{\infty} d\xi d_0(\xi)\left(1 + \ln(\rho / \rho_0)\frac{d}{3d\xi}(d_0(\xi))\right)\delta\left(\xi - \ln(T / T_0) + \frac{d_0(\xi)}{3}\ln(\rho / \rho_0)\right);$$

$$d(\rho = \rho_0;T) \equiv d_0(\ln(T / T_0)) \qquad (9)$$

At the same time, with respect to the behavior of the function $d_0(\ln T / T_0)$ in the limit of large values of the argument, the condition assumes its fairly rapid tendency to zero.

Taking into account Eq. (9), the EOS (3), (4) has the following form:

$$p(\rho;T) = \rho\left(\frac{kT}{m} + u_0\frac{d(\rho;T)}{3}\left(\frac{\rho}{\rho_0}\right)^{d(\rho;T)/3}\right);$$

$$u(\rho;T) = \frac{3kT}{2m} + u_0\left(\frac{\rho}{\rho_0}\right)^{d(\rho;T)/3}; u_0 = const \qquad (10)$$

The form of the function $d(\rho;T)$ in Eq. (10) is defined by the solution (9). From Eq. (9) follows a relatively weak logarithmic dependence of the exponent of the degree of potential energy on density and temperature. With relatively small compressions $\varepsilon \equiv \frac{1}{3}\ln\left(\frac{\rho}{\rho_0}\right) \ll 1$, when Eq. (9),

taking into account the properties of the delta-function follow the representation:

$$d(\varepsilon;x) = d_0(T)\left[1 - \varepsilon\frac{d}{dx}(d_0) + 2\varepsilon^2\left(2\left(\frac{d}{dx}d_0\right)^2 + d_0\left(\frac{d^2}{dx^2}d_0\right)\right) + O(\varepsilon^3)\right];$$

$$x = \ln\frac{T}{T_0}; \varepsilon = \frac{1}{3}\ln\frac{\rho}{\rho_0} \tag{11}$$

For the finite values of parameter $\varepsilon = \varepsilon(\rho)$, the function (14) in case $\frac{d}{dT}d_0 < 0$ can maintain

smoothness only under condition [32], [33]:

$$\frac{1}{3}\ln\left(\frac{\rho}{\rho_0}\right) \equiv \varepsilon < \varepsilon_{th} = \frac{1}{\max\limits_T\left(T\left|\frac{d}{dT}d_0(T)\right|\right)} \tag{12}$$

Condition (12) is similar to the criterion for the existence of a smooth solution of the Riemann equation only on a finite time interval, when the analogue of the time variable is a parameter $\varepsilon$ determined by the amount of compression of the medium. If condition (12) is violated, the value of the square of the derivative of the function (9) already has a singular behavior depending on the compression value. In fact, this may be an analogue of a phase transition of the second kind, when the discontinuity takes place not for the thermodynamic functions themselves, but for their

derivatives. Indeed, it follows from (9) that for the case $\frac{d}{dT}d_0(T) < 0$ only under condition (12)

the convergence of the integral having the form [32], [33] is ensured:

$$\int\limits_{-\infty}^{\infty}dx\left(\frac{\partial}{\partial x}d(x;\varepsilon)\right)^2 = \int\limits_{-\infty}^{\infty}dx\frac{\left(\frac{d}{dx}d_0(x)\right)^2}{1 + \varepsilon\frac{d}{dx}d_0(x)} < \infty \tag{13}$$

Let us establish a correspondence between the function $d_0(T)$ and the dependence of pressure on temperature at a fixed density. To do this, we use the representation for the pressure in (10) in the limit $\rho \to \rho_0$. At the same time, from (10) we obtain:

$$d_0(\ln(T/T_0)) = \frac{3}{u_0}\left(\frac{p(\rho = \rho_0;T)}{\rho_0} - \frac{kT}{m}\right) \tag{14}$$

To find the dependence of pressure on temperature for a given density value, one can use the known temperature dependence for the coefficient $\alpha(T;\rho) = \left(\frac{\partial\ln p}{\partial T}\right)_\rho$ (see Fig. 1):

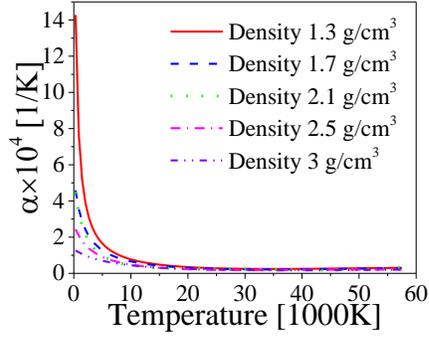

Fig. 1 The dependence of the coefficient $\alpha = \dfrac{1}{p}\left(\dfrac{\partial p}{\partial T}\right)_{\rho=const}$ on the temperature obtained in the simulation for water behind shock wave front [34].

As a result, without any additional assumptions, based only on the first principles used to obtain the EOS (10), it is also possible to obtain analytical representations for such important thermodynamic quantities as the specific isochoric heat capacity $c_V = \left(\dfrac{\partial u}{\partial T}\right)_\rho$, specific isobaric heat capacity $c_p = \left(\dfrac{\partial (u + p/\rho)}{\partial T}\right)_p$, isothermal compressibility $\beta_T = 1/\rho c_T^2$, $c_T^2 = \left(\dfrac{\partial p}{\partial \rho}\right)_T$ and the thermal expansion coefficient $\alpha_p = \beta_T \left(\dfrac{\partial p}{\partial T}\right)_\rho$.

A refinement can be carried out also for parameters such as the Boyle temperature and the critical point $(T_C; \rho_C; p_C)$ determined from conditions $\left(\dfrac{\partial p}{\partial \rho}\right)_T = \left(\dfrac{\partial^2 p}{\partial \rho^2}\right)_T = 0$ [5], [6] based on the obtained analytical form of the EOS.

A more detailed comparison with the results obtained from empirical EOS and calculations by the method of molecular dynamics [5], [6], [35], [36] can also be considered separately (see also the Subsection 3 in Appendix A).

### 3. The complete Mie-Grüneisen EOS for water

It follows from the solutions obtained above that the uniformity exponent of the function characterizing potential energy has a very weak logarithmic dependence on temperature and an even weaker doubly logarithmic dependence on density. This circumstance allows us to consider an approximation in which, at significant finite intervals of temperature and compression, a characteristic constant value of this parameter $d$, corresponding to these intervals can be used by analogy with the exponent of the isothermal EOS of water of the Taita-Muranheim type [2]. Here such an approximation is used to obtain the complete Mie-Grüneisen EOS.

It should be borne in mind that there is the following relationship between the internal potential energy $u_P(\rho;T)$ considered in (4) and (10) and the thermal part of the internal energy $u_T(\rho;T)$ (see Introduction), used in the representation of the EOS in the form of Mie-Grüneisen: $u_T = u_p - u_\rho + \frac{3kT}{2m}$ .

The Grüneisen coefficient in the Mie-Grüneisen EOS should be distinguished from the Grüneisen parameter $\Gamma = \frac{1}{\rho} \left( \frac{\partial u}{\partial p} \right)_{\rho=const}^{-1} = \frac{c_T^2 \beta}{c_V}$, which describes the compressibility of the medium and is a measure of thermal pressure, where $\beta = -\frac{1}{\rho} \left( \frac{\partial \rho}{\partial T} \right)_{p=const}$ [37]. For an isothermal EOS of the Tait–Murnaghan type, the pressure and compression are $p \propto \delta^n$ and $\delta = \rho/\rho_0$, where $n = \Gamma + 1$ [38] or $n = 2\Gamma + \frac{1}{3}$ relates the Grüneisen parameter $\Gamma$ and the exponent $n$ [39]. It was shown that, in the limit of high temperatures, the values of the Grüneisen parameter and the Grüneisen coefficient coincide [37]. For the special case $\partial c_V / \partial T = 0$, they also coincide with each other $\Gamma_T = \Gamma$.

Let us consider the common case when $\Gamma_T = \Gamma_T(\rho; T) \neq \Gamma$. We use the suggestion about parameter $d \approx const$ in (3), (4) and one obtains [1]:

$$u = \frac{3}{d} \frac{p}{\rho} + 3 \frac{kT}{m_{H_2O}} \left( \frac{1}{2} - \frac{1}{d} \right). \tag{15}$$

Using Eq. (5) and the Grüneisen relation $p_T = \rho \Gamma_T u_T$, one obtains

$$T \left( \frac{\partial (u_T \Gamma_T)}{\partial T} \right)_{\rho=const} = u_T \Gamma_T - \rho \left( \frac{\partial u_T}{\partial \rho} \right)_{T=const} . \tag{16}$$

In Eq. (15), for simplicity, the energy of the molecules rotation and the energy of intra-molecular vibrations are not taken into account.

Here, $p = p_\rho + p_T$ and equality $p_\rho = \rho^2 du_\rho / d\rho$ is used [17] [20]. Note, however, that in [20] the indicated relation between the functions $u_\rho(\rho)$ and $p_\rho(\rho)$ is obtained as a consequence of the first and second laws of thermodynamics only for the case of a constant value of the specific heat capacity at a constant volume, when the thermal part of the internal energy is a linear function of temperature. In case of an arbitrary dependence on temperature for the Grüneisen coefficient and the specific heat capacity, it is relation (16) that is a necessary condition for the thermodynamic consistency of the complete EOS, in the form generalizing the original Mie-Grüneisen EOS. At the same time (16) replaces condition $p_\rho = \rho^2 du_\rho / d\rho$, obtained in [20], which is no longer necessary, but for simplicity we will use it in addition to equation (16). Indeed, it is equation (16), following from relation (5), that generally ensures the fulfillment of the known necessary condition considered in [20], which any complete equation of state must satisfy in order to be consistent with the first and second laws of thermodynamics [7], [22].

Similarly, using Eq. (15), complementing Eq. (16), the equation for internal energy density can be obtained as follows

$$u_T - \frac{3}{|d|} u_T \Gamma_T = f(\rho) + A_T T . \tag{17}$$

Here, $f(\rho) = \frac{3}{d} \frac{p_\rho}{\rho} - u_\rho(\rho)$ and $A_T = \frac{3k}{m_{H_2O}} \left( \frac{1}{2} - \frac{1}{d} \right)$. The system of Eqs. (16) and (17) allows one to determine the EOS in the form of Mie-Grüneisen for an arbitrary medium. Let us note here that this approach assumes an isentropic or isothermal relation between the density and the temperature-independent part of the pressure. That is, in Eq. (17), the form of $f(\rho)$ is assumed later to correspond to the isentropic EOS of water in the Tait–Murnaghan form (see below Eq. (19)) [2], [38]-[41].

Now, using function $z(\rho\,;T) = u_T\Gamma_T$ and substituting this representation in Eq. (16), we obtain a partial differential equation:

$$\frac{\partial z}{\partial x} + \frac{3}{d}\frac{\partial z}{\partial y} - z = -f_1(y). \tag{18}$$

Here, $x = ln(\,T/T_0)$, $y = ln(\,\rho/\rho_0)$ and $f_1(y) = df(\rho(y))/dy$. Let us consider the solution of Eq. (18) for the case when in Eq. (17) functions $p_\rho(\rho)$ and $f_1$ have the forms

$$p_\rho - p_0 = \frac{\rho_0 c_0^2}{n}(\delta^n - 1)\,;\delta = \frac{\rho}{\rho_0}$$

$$f_1(y) = \frac{c_0^2}{n}\left(\frac{3(n-1)}{d} - 1\right)exp(y(n-1)) + \left(\frac{c_0^2}{n} - \frac{p_0}{\rho_0}\right)\left(\frac{3}{d} + 1\right)exp(-y). \tag{19}$$

Here, $c_0^2 = \left(\dfrac{\partial p}{\partial \rho}\right)_{0;T=T_0=const}$ is the square of the isothermal velocity of sound in undisturbed water and $n$ is the dimensionless exponent, which in general depends on temperature and pressure. This dimensionless exponent is commonly assumed constant when considering a rather narrow range of pressure. For example, $n \approx 7$ is used for pressures $\leq$25 kbar in [2], [38]-[41].

For the representations (19), the general solution of Eq. (18), has the form

$$z(x\,;y) = e^x F_0\left(y - \frac{d}{3}x\right) - \frac{c_0^2}{n}e^{y(n-1)} + \left(\frac{c_0^2}{n} - \frac{p_0}{\rho_0}\right)e^{-y}\ . \tag{20}$$

Here, $F_0$ is the arbitrary smooth function. Now, using representations (19) and Eq. (20), we obtain the EOS for water in the form of Mie-Grüneisen, i.e., $p_T = \delta\rho_0 z(\delta\,;T)$):

$$p = p_\rho + p_T = \delta\frac{\rho_0 T}{T_0}F_0\left(ln\left(\delta\left(\frac{T_0}{T}\right)^{\frac{3}{d}}\right)\right). \tag{21}$$

Here, $F_0(0) = \frac{p_0}{\rho_0}$ is the boundary condition and $T_0$ is the temperature of the undisturbed state of the medium characterized by the corresponding values of density $\rho_0$ and pressure $p_0$.

From Eq. (15) and Eq. (21) for the total internal energy density $u = u_\rho + u_T$, one obtains

$$u = \frac{3kT}{m_{H_2O}}\left(\frac{1}{2} - \frac{1}{d}\right) + \frac{3T}{dT_0}F_0\left(ln\left(\delta\left(\frac{T_0}{T}\right)^{\frac{3}{d}}\right)\right) \tag{22}$$

Thus, Eq. (21) and Eq. (22) are represents the self-consistent EOS in the form of Mie-Grüneisen, where the pressure and internal energy density depend on the density and temperature. However, to find these dependencies, function $F_0$ should be determined. Here, it is assumed that the parameters $n > 0$ and $d > 0$ for each considered range of compressions and temperatures are constant values.

To determine the function $F_0$ we use the Rankin–Hugoniot relation as a second boundary condition, using the conservation energy law at the shock-wave (SW) front [2]:

$$u - u_0 = \frac{1}{2}(p + p_0)\left(\frac{1}{\rho_0} - \frac{1}{\rho}\right). \tag{23}$$

By substituting of Eq. (21) and Eq. (22) in Eq. (23) one obtains

$$\frac{T}{T_0}F_0 = \frac{2d}{6 - d(\delta - 1)}\left[\frac{p_0}{\rho_0}\left(\frac{3}{d} + \frac{(\delta - 1)}{2\delta}\right) - \frac{3k(T - T_0)}{m_{H_2O}}\left(\frac{1}{2} - \frac{1}{d}\right)\right] \tag{24}$$

Now, using Eq. (20) and Eq. (23) and considering Eq. (24), we obtain explicit analytical representations for the thermal part of the internal energy density $u_T$ and for the Grüneisen coefficient $\Gamma_T$:

$$u_T = u - u_\rho = \frac{6}{6-d(\delta-1)}\left[\frac{p_0}{\rho_0}\left(\frac{3}{d} + \frac{(\delta-1)}{2\delta}\right) - \frac{3k(T-T_0)}{m_{H_2O}}\left(\frac{1}{2} - \frac{1}{d}\right)\right] + A \tag{25}$$

$$\Gamma_T = \frac{z}{u_T}; z = \frac{T}{T_0}F_0 - \delta^{n-1}\frac{c_0^2}{n} + \delta^{-1}\left(\frac{c_0^2}{n} - \frac{p_0}{\rho_0}\right) \tag{26}$$

Here, $A = \frac{3kT}{m_{H_2O}}\left(\frac{1}{2} - \frac{1}{d}\right) - u_\rho$ and $u_\rho = \delta^{n-1}\frac{c_0^2}{n(n-1)} + \delta^{-1}\left(\frac{c_0^2}{n} - \frac{p_0}{\rho_0}\right)$.

From Eq. (21) and Eq. (24) for the Hugoniot shock adiabatic, we obtain the following dependence of pressure $p$ and internal energy density $u$ on temperature and compression magnitude:

$$p = \frac{2\rho_0\delta d}{(d(\delta-1)-6)}\left[\frac{3k(T-T_0)}{m_{H_2O}}\left(\frac{1}{2} - \frac{1}{d}\right) - \frac{p_0}{\rho_0}\left(\frac{3}{d} + \frac{\delta-1}{2\delta}\right)\right] \tag{27}$$

$$u = \frac{6}{(d(\delta-1)-6)}\left[\frac{3k(T-T_0)}{m_{H_2O}}\left(\frac{1}{2} - \frac{1}{d}\right) - \frac{p_0}{\rho_0}\left(\frac{3}{d} + \frac{\delta-1}{2\delta}\right)\right] + \frac{3kT}{m_{H_2O}}\left(\frac{1}{2} - \frac{1}{d}\right) \tag{28}$$

Note that for the case of a polytropic ideal medium with adiabatic index $\gamma > 1$ for a shock adiabatic, a dependence of the pressure similar to (27) was obtained (see (89.1) in [42]):
$\frac{p}{p_0} - 1 = \frac{\delta-1}{(\gamma+1)/2\gamma - \delta(\gamma-1)/2\gamma}$.

For the dependence of temperature on pressure and compression on the Hugoniot shock adiabatic (27), we use the following representation (see also below Fig.2):

$$\frac{T}{T_0} = 1 + h\left(\frac{p}{p_0}\frac{(6-d(\delta-1))}{\delta(2-d)}\right). \tag{29}$$

To obtain (29), the estimation $h = (\frac{3kT_0\rho_0}{m_{H_2O}p_0})^{-1} \approx 0.000247$ is used, where $T_0 = 293 K^0$; $c_0 = 1483\frac{m}{s}$, $p_0 = 1bar$; $p^0/\rho_0 = 10^{-4}c_0^2/2.2$, $k = 1.38 \times 10^{-16}g \cdot cm^2/s^2 K^0$; $m_{H_2O} = 3 \times 10^{-23}g$. In (29), for the values $p$, $\delta$ and $T$ it is necessary to use the values according to the data presented in Table 3 in [43] to obtain the values of $|d|$ represented in Fig. 2.

Figure 2 shows the dependence of the temperature on pressure at various compressions and parameter $d$ values in (29), which is obtained from Eq. (21) using Eq. (24). The comparison of this dependence with known data [14], [43], [44] is presented in Fig. 2. In [35], the shockwave and thermodynamic data pertaining to the Hugoniot curve are tabulated in Table III as a function of shock pressure. The specific volume, shock velocity, particle velocity, internal energy density, and enthalpy in this table were obtained from the analytical fit of the experimental Hugoniot curve data for water and from the Rankin–Hugoniot relations (see (1)-(3) in [44]). The SESAME tabular form of the EOS of water is given in [14], which allows one to determine the temperature value at a given pressure, compression, and internal energy density.

As a result, Fig. 2 shows the temperature-pressure dependence curve corresponding to the data in Table 3 of [43] and the parameter $|d|$ values versus the pressure at which an exact correspondence between data [43] and equation (29) is realized. Figure 2 also shows the temperature-pressure dependence obtained using the SESAME EOS of water [14]. One can see a satisfactory correspondence between the curve following from the EOS for water near the shock adiabatic obtained in this paper and the tabular form of this SESAME EOS.

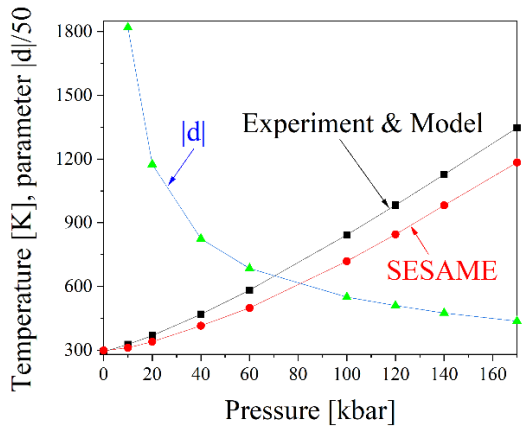

Fig. 2 The curve with black squares denotes the dependence of temperature on pressure, which corresponds to the exact solution (29), as well as to experimental data given in Table 3 [43], when using in (29) the values of the intermolecular interaction potential exponent $|d| = d > 0$ corresponding to the curve indicated by green triangles. For comparison, we present the dependence of temperature on pressure denoted by red circles obtained using the SESAME EOS of water [14].

The Table 1 below shows the values of the parameters corresponding to the data given in Fig. 2 on the magnitude of pressure $p$, compression $\delta = \rho / \rho_0$, temperature $T$, as well as the velocity $D$ of the shock wave and the velocity $U$ of the medium behind the shock wave front in water according to Table 3 of [43].

Table 1

| $p(Kbar)$ | $\delta$ | $T^0 K$ | $D(km/\sec)$ | $U(km/\sec)$ | $d$ |
|---|---|---|---|---|---|
| $p_0 = 10^{-3}$ | 0.9982 | 293 | 1.483 | 0 | - |
| 10 | 1.2189 | 327 | 2.352 | 0.426 | 36.411 |
| 20 | 1.3187 | 369 | 2.871 | 0.698 | 23.497 |
| 40 | 1.4409 | 469 | 3.611 | 1.110 | 16.487 |
| 60 | 1.5242 | 583 | 4.173 | 1.440 | 13.722 |
| 100 | 1.6461 | 843 | 5.045 | 1.986 | 11.019 |
| 120 | 1.6969 | 983 | 5.404 | 2.225 | 10.195 |
| 140 | 1.7422 | 1127 | 5.731 | 2.447 | 9.541 |
| 170 | 1.8057 | 1347 | 6.160 | 2.760 | 8.7413 |

The parameter $d$ characterizes the paired effective interaction potentials. This coefficient becomes useful in the analysis of experimental data for strong shock wave propagation in water at pressures ≥80 Kbar [45]. In this range of pressures, there is a simplification of molecular interactions, which is rather complicated at lower temperatures and pressures [45]. For example, the repulsive potential of the Lennard–Jones type with an exponent $d = 12$ is also considered in [45] in accordance with Fig. 2 for pressures >60 Kbar.

Thus, the exact coincidence on Fig.2 between the analytical EOS (29) and the observational data from Table 3 [43] is established due to the appropriate choice of the only one fitting parameter $d$ . For example, at the pressure 80 kbar and temperature 600 K an exponent value is equal $d = 12$ in accordance with [45].

### 4. The Grüneisen parameter and the isothermal EOS

In the limit $\delta^{n-1} >> 1$ of Eq. (26), the Grüneisen coefficient can be approximated as

$$\Gamma_T = n - 1. \tag{30}$$

Note that the relation between the exponent $n$ and the Grüneisen parameter $\Gamma$, obtained in [38], exactly coincides with (30) if $\Gamma_T = \Gamma$. Thus, in this approximation the Grüneisen coefficient and the Grüneisen parameter are equal to each other $\Gamma_T = \Gamma$ . Relation (30) is also known for the case of an ideal gas if the exponent $n$ coincides with the adiabatic exponent $n = \gamma$ [2]. Indeed, EOS (19) is similar to the EOS of an ideal polytropic medium [42], where the adiabatic index $\gamma = c_p/c_V$ plays a role similar to the exponent $n$ in (19).

In the limit $\delta \rightarrow 1 + 6/d$ at arbitrary temperatures (or in the limit of large temperatures at arbitrary compressions) using Eq. (26), the Grüneisen coefficient $\Gamma_T$ reads

$$\Gamma_T = \frac{d}{3} \quad . \tag{31}$$

Thus, from Eq. (30) and Eq. (31) the relation $n = 1 + d/3$ is valid in the limits $\delta^{n-1} >> 1$ and $\delta \rightarrow 1 + 6/d$.

Let us consider an arbitrary value of $d$ as the fitting parameter in the relationship between the exponent $n$ in the EOS of water (19) and the Grüneisen parameter.

The Grüneisen parameter $\Gamma = c_T^2 \beta/c_V$ has the general form (see Appendix C, the representations for $c_T^2$ in (C9) and (C2) for $c_V$)

$$\Gamma = \frac{d}{3\left(1 + \frac{dk\delta^{-n+1}}{m_{H_2O}c_0^2\beta(T;p)}\left(\frac{1}{2} - \frac{1}{d}\right)\right)} \quad . \tag{32}$$

The function $\beta = \beta(T;p)$ in (32) is shown in Fig. 3. As noted above, at sufficiently high temperatures, one obtains $\Gamma_T = \Gamma$. From (32) in the limit $\delta^{n-1} >> k/c_0^2 m_{H_2O}\beta$, it follows that $\Gamma \rightarrow d/3$. The latter coincides with the estimate (30) for the Grüneisen coefficient.

For the case $d \neq 2$ ; $\Gamma \neq \frac{d}{3}$ in (32) it is possible to obtain a relation of $n$ with compression, temperature, and pressure (see Fig. 3 for the dependence of $\beta(T;p)$ on pressure and temperature):

$$n = 1 + \frac{1}{\ln \delta} \ln \left[\frac{3A_T\Gamma(d-2)}{2(d-3\Gamma)}\right] \tag{33}$$

Here, $A_T = \frac{k}{m_{H_2O}c_0^2\beta(T;p)}$.

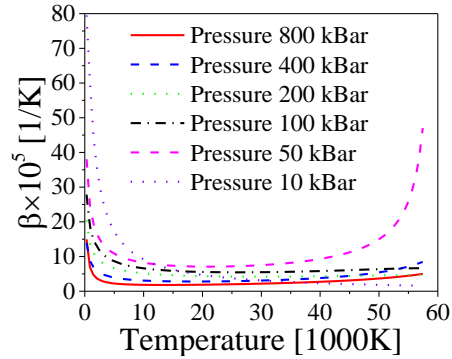

Fig. 3. Dependence of the volume expansion coefficient $\beta = -\frac{1}{\rho}\left(\frac{\partial \rho}{\partial T}\right)_{p=const}$ on temperature for water behind the shock wave front, obtained in the simulation [34].

Let us apply Eq. (33) for a given compression value; the exponent of the potential of intermolecular interaction in water reads

$$d = \frac{6\Gamma(1 - A_T \delta^{-(n-1)})}{2 - 3A_T \Gamma \delta^{-(n-1)}} \qquad (34)$$

Now, using the data presented in [43] (see Table 3 in [43]) for the local sound velocity $c_T$ at different compressions and Eq. (C9), one can calculate the values of $n$ at different compressions.

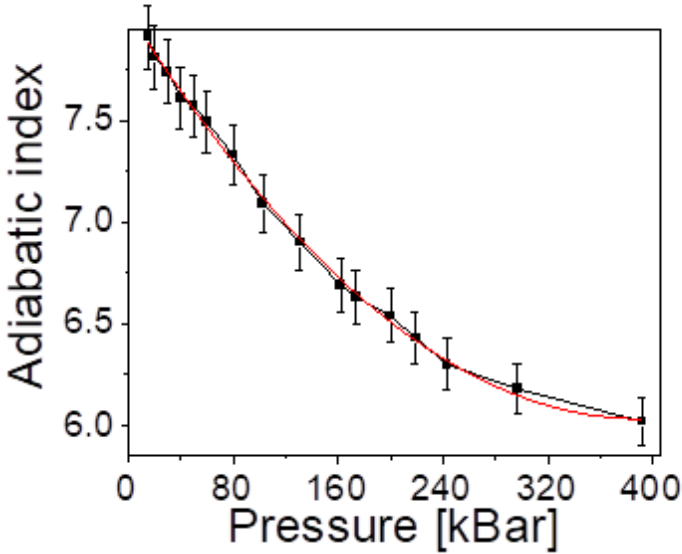

Fig. 4 Dependence of the exponent $n$ for isothermal EOS on pressure for water behind the shock wave front, obtained in the experiment and simulation [34].

The Table 2 below shows the values of the adiabatic index or an exponent $n$ for the isothermal EOS corresponding to the data (black points) given in Fig. 4 on the magnitude of pressure $p$ or compression $\delta = \rho / \rho_0$ behind the shock wave front in water and according to relation $n = 1 + 2\ln(c_T(p)/c_{0T}(p_0))$, obtained from (C9) (see Appendix C), where the isothermal speed of sound behind the shock wave front from the Table 3 in [43].

Table 2

| $p(kBar)/\delta$ | 10/1.22 | 20/1.32 | 40/1.44 | 60/1.52 | 100/1.65 | 120/1.697 | 140/1.74 | 170/1.8 |
|---|---|---|---|---|---|---|---|---|
| $n$ - [43] | 7.42 | 6.9 | 6.44 | 6.29 | 5.89 | 5.76 | 5.65 | 5.52 |
| $n$ - Fig.4 | 7.95 | 7.75 | 7.63 | 7.5 | 7.15 | 6.9 | 6.75 | 6.6 |

Using the temperature dependence of the coefficient $\beta(T)$ (see Fig. 3) and Eq. (34), the estimates of the modulus of the potential of intermolecular interaction $d$ in water can be obtained. For example, for $\delta \approx 1.32$, according to [46] (see function $\Gamma = \Gamma(\delta)$ in Fig. 11 in [46]) and [43] ($c \approx 3390 m/s$; $c_0 \approx 1483 m/s$) we estimate $n \approx 6.9$; $\Gamma \approx 0.9$ (see Table 2). For such compression, as follows from [35], the pressure is $p \approx 20$ kbar and the temperature is $T = 96^0 C$. In this case, the coefficient of thermal expansion is equal to $\beta \approx 0.0008 \text{grad}^{-1}$ and $A_T \approx 0.29$. As a result, we estimate $|d| \approx 2.76$. Thus, according to (34), there is a non-monotonic dependence of $d$ with increasing compression. However, according to Fig. 4 and Table 2, there is a monotonous decrease in the exponent $n$ versus the increase in compression in the isothermal EOS of water (19). Indeed, according to Eq. (33) for a fixed value of $d$, a decrease in the Grüneisen parameter (see Fig. 5 and Table 3 below) with an increase in compression from 1.6 to 2.3 may lead to a decrease in the value of $n$ in the isothermal EOS of water (19).

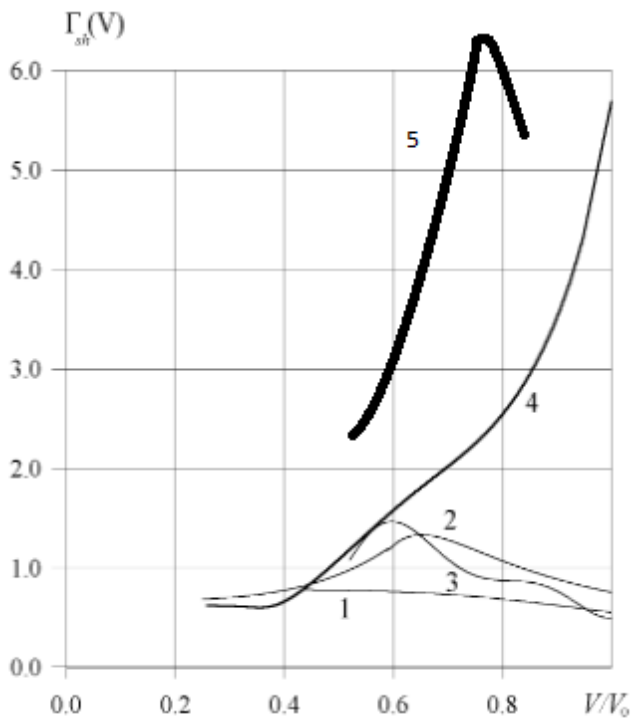

Fig.5

Fig. 5 is taken from [50] (see Fig.6 in [50]). The dependences 1-4 of the Grüneisen parameter $\Gamma = \Gamma_{sh}(V/V_0); V/V_0 = 1/\delta$ for water behind the shock wave front are obtained from: 1- [47], 2-[48], 3-[49], 4- [50] and 5- from Eq. (32) and Table 3.

Table 3

| $p(kBar)$ | 10 | 20 | 40 | 60 | 100 | 140 | 170 |
|---|---|---|---|---|---|---|---|
| $T^0 K$ | 327 | 369 | 469 | 583 | 843 | 1127 | 1347 |
| $\delta^{-1} = V/V_0$ | 0.82 | 0.75 | 0.69 | 0.66 | 0.61 | 0.57 | 0.55 |
| $d$ | 36.4 | 23.5 | 16.5 | 13.7 | 11.02 | 9.5 | 8.7 |
| $\Gamma$ | 5.27 | 6.45 | 3.89 | 3.38 | 2.89 | 2.63 | 2.48 |

The data in Table 5 correspond to Curve 5 in Fig.5 and are obtained on the basis of formula (32) and parameter values $d, n, \beta$, presented in Table 1, Table 2 (rows corresponding to data [43]) and Fig.3, respectively.

Note that curve 5 in Figure 5 has a non-monotonic dependence of the Grüneisen parameter on the compression value, which is characteristic of fluids. This qualitatively corresponds to the behavior of curves 1, 2, and 3 which were obtained at fixed values of temperature. A noticeable quantitative difference between 5 and the other curves in Figure 5 may be due to a dependence on temperature for the parameters that determine the values of the Grüneisen parameter described by formula (32) and Table 3.

The indicated difference between curve 5 and curve 1 takes place despite the correspondence between the formula for pressure (21) obtained in our paper and the formula (1.11) in [47], given without derivation. There is also a correspondence between the representation for pressure in formula (27) and the formula (1.1) obtained in [47] from empirical considerations. At the same time, the reason for the difference between curve 5 in the Fig.5 and the other curves is precisely the presence of a dependence on compression and temperature for the exponent $d$ in (32) and Tables 1 and 3, which is not considered in [47]. Indeed, in [47], a relation $n - 1 = d/3$ is used, which, with the empirically established values of $d$ in the range from $d = 9$ to $d = 12$ given there, leads to the corresponding possible range of values of the exponent from $n = 4$ to $n = 5$.

Let us note that a relation $n = 1 + d/3$ is also obtained before from Eq. (30) and Eq. (31) (see after (31) and also near the same Eq. (6.7) in [51]).

If, on the other hand, another exponent value is used at a temperature of 448 K, considered in [47], $n = 6.44$ is used, obtained according to Table 2 at a close temperature of 469 K, then a value $d = 16.32$ is obtained, which is already close to the value 16.5 given in Table 3.

As indicated in [47], an EOS for water considered for compressions $2.3 > \delta > 1$, can be applied not only at temperatures below 448 K, but also for higher temperatures, at which, however, dissociation and ionization processes are not essential.

However, unlike the conclusions of [47], for the also nowadays important (see [51]-[53]) determination of the Grüneisen parameter, the representations obtained in our work for the EOS do not depend on the model estimation of the specific heat capacity $c_V$. Therefore, the restrictions imposed on the scope of applicability in [47]-[50] are not mandatory for the EOS considered in our paper. Instead, the essential constraints are the conditions imposed on the parameter $d$ in the modification of the general form of the Mi-Grüneisen type EOS discussed above. For example, from Eq. (27), the estimation of the specific heat capacity $c_V$ in Eq. (C2) gives the substance dependence on the exponent $d$ in the form $c_V = 3k(d-2)\big(d(\delta-1)-4\big)/2m_{H_2O}d\big(d(\delta-1)-6\big)$. In Eq. (2.2) of [47] the dependence of $c_V$ on the density and temperature is also obtained on the base of Eq. (15) and (21), but for the different representation of an arbitrary function $F_0\left(\ln\delta + \dfrac{3}{d}\ln\left(\dfrac{T_0}{T}\right)\right)$ in (21). Moreover the argument of function $F_0$ is function of the unknown function $d = d(\delta;T)$ and independent variables that are existed in the Riemann-Hopf equation (7).

Thus, it seems most promising to consider the EOS in an alternative form that automatically takes into account the necessary restrictions on the parameter $d$, which follows from the first principles and lead to a Riemann-Hopf equation for this parameter, which depends on density and temperature (see Chapter 2).

# 5. Discussion

Thus, a wide-range EOS of an arbitrary medium is obtained from the first principles, as well as a complete EOS of water in the form of Mie-Grüneisen. The results obtained are important both in connection with the problem of closing the equations of hydrodynamics of a compressible medium and for understanding of the thermodynamic properties of pure fluids and there mixtures up to very high temperatures and pressures.

For example, the last are required in geophysics and astrophysics because very few experimental $p - \rho - T$ data exist at these extreme conditions because water is thought to be one of the most abundant compounds in ice giants such as Neptune and Uranus [54]. Therefore EOS for water is critical to understanding there temperature and pressure distributions when the ANEOS, SESAME models [14], [55] and Quantum Molecular Dynamics (QMD) calculations [56] and QMD based EOS model [57] have been employed for modeling interior of ice giants as in the shock experiments [54], [58]-[61].

The EOS known to date take into account the effects associated with the interaction of medium particles only in a certain limited range of changes in pressure, density and temperature [1], [2], [5], [6]. This limitation is due to a number of assumptions used in the derivation of the EOS. In this paper, a wide-range EOS of an arbitrary medium is obtained from the first principles, in which the interaction between the particles of the medium is taken into account without involving any additional assumptions. This makes it possible to assess the degree of adequacy of various approximations used to describe the medium when taking into account the interaction between the particles of the medium. For example, there is an idea of the applicability of the known Tate-Muranheim type isothermal EOS of water (19) with the value of the exponent $n = 7.15$ in the pressure range up to 25 Kbar [2], [38]-[41]. As represented in (33) and (34) with an increase in compression and pressure, this exponent can change, however, remaining almost unchanged in the same limited ranges of pressure changes, as well as the associated exponent of the power-law for the interaction potential between particles in the imperfect media.

At the same time, the obtained exact solution for the EOS, taking into account the dependence of the potential energy of the interaction of particles on the temperature of the medium and its density, opens up new opportunities for the analysis of traditional approaches in statistical mechanics of both thermo-dynamical equilibrium and non-equilibrium systems.

Indeed, all books on statistical mechanics have a chapter devoted to imperfect gases, in which the virial expansion of the EOS in terms of density is derived [1]. Up to now an explicit evaluation of the second virial coefficient is given only in simple cases as hard spheres and square well potentials [5], [6], [62]-[67]. For example, the Boyle temperature $T = T_B$ can be set at which the second virial coefficient vanishes $B(T = T_B) = 0$ (when $B > 0$ for values $T > T_B$), as well as the temperature $T = T_{max}$ at which the coefficient $B$ reaches its maximum value, as a function of temperature $(\partial B / \partial T)_\rho = 0; (\partial^2 B / \partial T^2)_\rho < 0$.

The value of coefficient $B$ may be obtained from the EOS by using known relation $B = \lim_{\rho \to 0} \left( \frac{\partial (pm / kT\rho)}{\partial \rho} \right)_T$ [5], [6]. Using this definition of the second virial coefficient in Appendix A, an explicit analytical representation (A 21) is obtained for this coefficient from equation of state (A19). From (A 21), in particular, it follows that the finite (nonzero or not infinite) value of the coefficient $B$ is not always feasible in the limit of arbitrarily small density corresponding to the specified definition of this coefficient. This indicates the possibility of limitations in the application of the virial decomposition method not only at high densities, when, as is known, the corresponding series on the degrees of density weakly converge, but also vice versa, in the limit of small densities.

Moreover, usually, only numerical methods are used to estimate this coefficient. However, a simple analytical closed form is also obtained before yet in [62], which is especially useful at low temperatures where the series expansion of virial coefficient converges very slowly. To estimate the second virial coefficient in [62], an alternative definition was used to the one given and used in [5], [6], which has the form $\widetilde{B}(T) = -\frac{2\pi}{3kT} \int_0^\infty dr r^3 \frac{dU}{dr} \exp\left( -\frac{U}{kT} \right)$ for the case of the Lennard-Jones potential $U = U_{LD} = 4\varepsilon_{LD} \left[ \left( \frac{\sigma}{r} \right)^{12} - \left( \frac{\sigma}{r} \right)^6 \right]$, describing the interaction between the particles of the medium depending on the distance between the particles $r$. For example, $\varepsilon_{LD} / k = 120 K^0; \sigma \approx 3.2 A$ in case of argon [64]. In particular, in [62], after the introduction of dimensionless variables $B^*(T^*) = \widetilde{B} / \left( 2\pi\sigma^3 / 3 \right); T^* = kT / \varepsilon_{LD}$ the following estimates were obtained for the Boyle temperature $T_B^* \approx 3.417927$ at which $B^*(T_B^*) = 0$ as well as the temperature $T_{max}^* \approx 25.15257$ at which the second virial coefficient reaches the maximum value. The definition of the second virial coefficient in [62] is considered as equivalent to the definition, which is based on the use of the EOS of the medium relating pressure, density, temperature [5], [6].

The physical meaning of the Boyle temperature corresponds to the temperature at which the forces of attraction and repulsion between the particles balance each other and the particles behave as in an ideal gas without interacting with each other.

At the same time, the question arises whether the exact analytical solution for the equation of state obtained in this paper makes it possible to determine the Boyle temperature if the assumption of homogeneity for the inter-particles interaction potential is used to obtain this solution? After all, the uniformity condition excludes the possibility of considering a superposition of repulsive and attractive potentials, as for the Lennard-Jones potential. Indeed, condition (1) determines the relationship between the average distance between the particles at a given compression value with the potential energy of interaction between the particles corresponding to either repulsion at a positive value of the constant coefficient $u_0 > 0$ in (1), or attraction at a negative value of this value. At the same time, there is no possibility of considering a linear superposition of attraction and repulsion potentials, since in this case the condition of uniformity will be violated, which allows the use of the form of representation of the virial theorem considered in this paper. However, the effect of mutual compensation of the forces of attraction and repulsion characteristic of the Boyle temperature can be obtained due to the fact that at this temperature the limit $d(T \to T_B) \to 0$ is realized.

In particular, the analog of the Boyle temperature can be determined from the solution of the corresponding transcendental equation of the form $T_B = (p(T_0)m/k\rho_0)\exp\int\limits_{T_0}^{T_B} dT\alpha(T)$, obtained from relation (14) by equating the right part (14) to zero, where the dependence of the function $\alpha(T)$ is shown in Fig. 1. From Eq. (34) in case $\Gamma \neq 2/3$, condition $d(T = T_B) = 0$ also gives the possibility to obtain the analog of the Boyle temperature from the solution of equation $A_T(T_B) = \delta^{n-1}$ (see Fig. 3 for the dependence of $A_T(T)$).

At a zero value of the power-law exponent $d = 0$, the potential of the interaction energy between the particles becomes, although not zero, but already constant, independent of the distance between the particles which gives a zero inter-particles force, as for an ideal gas. At the same time, not for any exact solution obtained in this paper, it is possible to implement such a condition for an exponent $d$, to vanishes, as, for example, is the case for the solution (A19) given in Appendix A. Although for this case it is possible to determine the magnitude (A21) for the second virial coefficient $B$, according to the representation obtained from the form of the EOS (A19) (see Appendix A). At the same time, according to the exact solution obtained in (A21), the dependence of the second virial coefficient on temperature makes it possible to determine not only the Boyle temperature (A22), but also the temperature values at which both the maximum and the local minimum are reached according to (A23). Until now, such a possibility of the existence of conditions under which the second virial coefficient has not only a maximum, but also a local minimum value at the different temperatures has not been considered. However, the permissibility of the existence of such an additional extremum of the second virial coefficient on temperature may indirectly follow, for example, from the data [68] (see blue points on Fig. 2-D in [68]) on the dependence of the second virial coefficient of the protein solution on the ionic strength of the salt solution, which may similarly influences on the interaction of particles as temperature.

Also an example of the possible application of the obtained in (9), (10) (or (35), (36)) new EOS is tied with the compare it to the well-known Carnahan-Starling (CS) EOS $pm/\rho kT = (1 + \eta + \eta^2 - \eta^3)/(1 - \eta)^3; \rho = \tilde{n}m$, where $\eta = 4\pi R^3 \tilde{n}/3$ is a packing fractions of particles of the same hard-core radius $R$ and $\tilde{n}$ is their particle density [46], [54]-[56]. The CS EOS was successfully applied in the theory of simple liquids [63], [64]. Also for the compressible nuclear liquid to obtain high-quality fit (of the ALICE hadron multiplicities measured at the center-of-mass energy 2.76 TeV per nucleon) when the modification of CS EOS with induced surface tension is used [65], [66]. The high accuracy of CS EOS is provided by the correct values of the first seven virial coefficients of the gas of hard spheres which are reproduced by this model. This EOS is more accurate at high densities than excluded-volume model, which is the same as the Van der Waals EOS with repulsion. It is important to note that the temperature dependence of pressure in the new self-consistent analytical form of EOS (9), (10) or even in a more simple EOS (27) for an arbitrary media has more complex and common form than in the usually used modifications of CS EOS. In case when $\delta = \rho / \rho_0 = \eta < 1$ the compare of EOS (27) in the high temperature limit ($T >> T_0; Tk/m >> p_0/\rho_0$) with CS EOS gives the possibility to obtain the estimation for the negative value of the effective power-law exponent $d$ in the following form:

$$d = -\frac{12\eta(2-\eta)}{(1-\eta)(4-5\eta+4\eta^2-\eta^3)}; \eta < 1 \qquad (35)$$

For example, in cases $\eta = 0.2; 0.22$ considered in [66] from (35) it is possible to obtain corresponding estimations $d \approx -1.71; -1.95$. Thus, according to (2) and (35), the potential energy of interaction between particles tends to zero when the average distance between particles decreases, which is similar to the well-known effect of asymptotic freedom in the theory of elementary particles [29], [30]. In the limit of arbitrarily dense packing $\eta \rightarrow 1$, the absolute value of the negative exponent in (35) also increases indefinitely, enhancing the effect of asymptotic freedom, for which the considered limit of high temperatures in (27) is not associated with the need to take into account quantum effects [64].

   We shall also focus on the question of using the EOS of a compressible medium to obtain close form for the hydrodynamics equations.

   Indeed, even when describing obviously non-equilibrium macroscopic processes using equations of hydrodynamics of a compressible medium, the EOS is used to obtain a closed description of the system under the assumption of the realization of local thermodynamic equilibrium [69], [2], [3]. L. Euler was the first who noted such a necessity of using the equation connecting pressure and density to close the hydrodynamic description in case of consideration of the dynamics of a compressible medium [69]: "Let us now turn to those elements which contain that which is unknown. In order to properly understand the motion that will be imparted to the fluid it is necessary to determine, for each instant and for each point, both the motion and the pressure [pression] of the fluid situated there. And if the fluid is compressible, it is also necessary to determine the density, knowing the above-mentioned other property which, together with the density, makes it possible to determine the elasticity. The latter, being counterbalanced by the fluid pressure, must be considered equal to that pressure, exactly as in case of equilibrium, where I have developed these ideas more thoroughly". Thus, L. Euler introduced a qualitative difference between the pressure included in the equation of state of the medium, which he called elasticity and dynamic pressure, the gradient of which is included in the Euler equation, although at the same time an assumption was made about the possibility of quantitative identification of these quantities, like in case of the medium is in an equilibrium state [69].

   How to justify such an assumption is the subject of further consideration. Indeed, even in case of a relatively weak disturbance of the equilibrium state in the presence of medium flows at speeds much lower than the speed of sound, when the incompressible medium approximation is valid, there is a dependence of the dynamic pressure on the gradients of the medium velocity. In this case, the closure of the hydrodynamic equations dispenses with the use of the EOS of the medium. Therefore, it does not seem quite consistent when, for obviously non-equilibrium near sonic and supersonic flows, the pressure is set only on the basis of an EOS that does not contain velocity gradients of a compressible medium in any explicit form generalizing the case of an incompressible medium. In this regard, in [70],[71] various variants of the closure of equations of hydrodynamics are considered in cases where it is possible to exclude the force associated with the pressure gradient by establishing its balance with some volumetric force acting on the medium. At the same time, for example, in [33], [71], the magnitude of the dynamic pressure is already explicitly determined $p = \varsigma div\vec{u}$ through the magnitude of the divergence of the velocity field of the compressible medium, taking into account the volumetric viscosity of this medium. Similar representations for pressure in case of a medium far from equilibrium, in which the volume (or second) viscosity coefficient can be much higher than the coefficient of ordinary shear viscosity, are also given in [2], [42]. Moreover, in [33], [71] it is shown that the expression (79.1) in [42] for the rate of dissipation of the integral kinetic energy of the compressible medium is in contradiction with the equations of hydrodynamics and this is a direct consequence of the use in [42] for the pressure gradient of the representation following from the equilibrium thermodynamic relation, which includes pressure having a meaningful of elasticity as in [69]. Appendix B summarizes this proof obtained in [33], [71] and indicates the need for a more balanced use of the EOS of the medium in this connection.

## Summary

An explicit complete analytical EOS for an arbitrary condensed media and EOS for water in the Mie-Grüneisen form are obtained. The last is presented as a dependence of pressure, the thermal part of the internal energy density, and the Grüneisen coefficient on water density and temperature. It is shown that a good correspondence between the obtained new analytical equation of state with observational data and the well-known tabular form of the EOS for water is achieved by taking into account the dependence of the effective exponent of the degree of potential of intermolecular interaction on the degree of compression. This allows one to use the ratio between pressure and internal energy to obtain an explicit form of the shock adiabatic. The latter is important for determining the explicit dependence of the Dyakov parameter, which is a key parameter, together with the Mach number when solving problems of shockwave stability [10]-[12]. It is shown that when using the isothermal EOS (19) to describe strong shock waves in water, it is necessary to consider the dependence of the exponent on the compression and temperature [72], [73]. Indeed, the results presented in [72], [73] show the importance of this dependence for the interpretation of the non-monotonic dependence of the converging strong shockwave velocity versus the shock radius observed experimentally. The relation was also obtained between the exponent and the Grüneisen parameter. It was shown that this relation significantly depends on the type of intermolecular interaction potential, which in its turn varies according to the water compression. An accurate analytical solutions, obtained on the basis of first principles for the dependence of the degree of interaction potential on temperature and density, may also be useful in connection with the machine-learning interatomic potential models that allows for model materials and allows for calculation of forces and molecular dynamics simulations [74].

## Acknowledgement

I am thanks Yakov Krasik for drawing my attention to this problem, and for his support, fruitful discussions, and kindly assistance in the manuscript editing. I also express my thanks to Alexander Virozub for providing simulation data and to Alexander Chefranov for his support. This research was supported by the Israeli Science Foundation Grant No. 492/18

### Appendix A. Virial theorem generalization and EOS

1. **Modification of Feynman's et al. theory [28].** Let us generalize the conclusion of the virial theorem proposed in [28] to the case of an arbitrary exponent $\tilde{d} = \tilde{d}(r)$ of the power-law of potential energy of interaction between pare of particles located at distances from each other $r$:

$$E_P \propto \frac{e^2}{r^{\tilde{d}}} \qquad (A1)$$

In (A1), as in [28], $e$ - is the charge of an electron, but unlike [28], where a Coulomb-type potential with a fixed value $\tilde{d} = 1$ is considered, here and further it will be assumed that the value $\tilde{d} = \tilde{d}(r)$ can be a variable function, and not a constant value. At the same time, we will assume that the function tends to zero in the limit of small distances (which is similar to asymptotic freedom in elementary particle physics [30]), as well as in the limit of large distances:

$$\tilde{d} \to 0, r \to 0 \qquad (A2)$$

$$\tilde{d} \to 0, r \to \infty \qquad (A3)$$

We will use the equation of balance when the change of total energy is caused only by changes in energy at the boundaries of the system. In this case, the outer boundary of the system is determined by its volume, and the inner boundary corresponds in (A 1) to the limit $r \to 0$ taking into account condition (A2).

For the first law of thermodynamics, we have a representation in the form of an energy balance equation in which the total energy of the system, represented as the sum of kinetic and potential energies $E = E_K + E_P$:

$$dE = TdS - pdV \qquad (A4)$$

Let us consider a similarity transformation in which all charges including the elementary charge $e$ are changed by the factor $\lambda_\varepsilon = 1 + \varepsilon, \varepsilon << 1; e \to \lambda_\varepsilon e$, all distances by the factor $\lambda_\mu = 1 + \mu, \mu << 1; r \to \lambda_\mu r$ and all the energies by $\lambda_\eta = 1 + \eta, \eta << 1; E \to \lambda_\eta E$ [28]. As in [28] the quantum of action $\hbar$ and the electron mass are assumed to be unchanged.

From the expression of the potential energy (A1) the following relation obtains:

$$\frac{e^2}{r^{\tilde{d}(r)}}(1+\eta) = \frac{e^2(1+\varepsilon)^2}{\exp\left[\tilde{d}\left(r(1+\mu)\right)\ln\left(r(1+\mu)\right)\right]} \approx \frac{e^2(1+2\varepsilon)}{r^{\tilde{d}(r)}\left[1 + \mu\left(\tilde{d} + r\ln r \frac{d}{dr}(\tilde{d}(r))\right) + O(\mu^2)\right]} \text{ or:}$$

$$\eta = 2\varepsilon - \mu d - O(\mu^2),$$

$$d \equiv \left(\tilde{d} + \frac{d}{dz}\tilde{d}(z)\right); z \equiv \ln(\ln r) \qquad (A5)$$

Relation (A 5) for the value $\tilde{d} = 1$ exactly coincides with the conclusion [28], and for the function $\tilde{d}(r)$ gives its generalization.

To obtain (A 5), it is taken into account that the potential energy (A 1) is considered outside the inner boundary at finite values $r$ when the function $\tilde{d}(r)$ is also finite. At the same time, (A 5) takes into account only the terms of the first order of smallness in terms of parameters $\varepsilon; \mu; \eta$. that determine the similarity transformation. Therefore, when converting to (A 5), only the first term of the specified Taylor series expansion $\tilde{d}(r(1+\mu)) \approx \tilde{d}(r) + r\mu\partial\tilde{d}(r)/\partial r + (\mu^2)$ in degrees of $\mu$ for the function $\tilde{d}(r(1+\mu))$ is taken into account.

The de Broglie wave-length, as all lengths, must change as $1 + \mu$ and thus momenta change as $(1+\mu)^{-1}$ and kinetic energies as $E_K \rightarrow E_K(1+\mu)^{-2} \approx E_K(1 - 2\mu + O(\mu^2))$[28]. However, these energies like all energies must change as $E_K \rightarrow E_K(1+\eta)$, hence from this and from (A5):

$$\eta = -2\mu \qquad (A6)$$

$$2\varepsilon = \mu(d-2) \qquad (A7)$$

In particular, for the value $d_1 = d = 1$ the ratio (A 7) coincides with the formula (31) of [28], and the ratio (A 6) coincides with the formula (30) of [28] for any values $d$.

Let us treat, as in [28], the effect of the change of charges by a perturbation treatment when the energy change introduced by changing charge. In this case, the change in the total energy of the entire system $\eta(E_K + E_P)$ should be equal to the sum of the changes caused by the change in charge at the inner boundary and outer boundaries. At the inner boundary, taking into account condition (A2) in (A 5), only a change in charge leads to a change in potential energy, which as a result is equal to $2\varepsilon E_P$. In order to take into account the change in energy at the outer boundary of the system under consideration, which has an arbitrary finite volume, we use the consideration proposed in [28] in the following form: "In order to arrive at the same configuration reached by similarity transformation, the volume must now be readjusted. This is done by a volume increase $(1+\mu)^3 \approx 1 + 3\mu$". Thus, energy of the system will decrease by amount equal $3\mu pV$, where $p$ is the pressure on the external boundary of that volume and the energy balance equation is obtained in the following form:

$$\eta(E_K + E_p) = 2\varepsilon E_P - 3\mu pV \qquad (A8)$$

After substituting in (A8) equalities (A6) and (A7) in form of $\mu = 2\varepsilon/(d-2); \eta = -4\varepsilon/(d-2)$ and carrying out the corresponding simplifying transformations, we obtain a known form of representation of the virial theorem in case when the potential energy of the interaction of particles is a homogeneous function of degree $n = -d$ [1]:

$$pV = \frac{2}{3}E_K + d\frac{E_P}{3} \qquad (A9)$$

In contrast to the well-known conclusion of the virial theorem (A9) given in [1], in the representation obtained in (A9), the value $d$ can already be an arbitrary function satisfying (A5) under condition (A2). In case when this value is $d = 1$ the ratio (A9) exactly coincides with the one derived in [28] (see (29) in [28]).

Thus, the ratio (A9) is obtained only on the basis of similarity transformations and the energy balance equation (A4). At the same time, the assumption about time averaging, which is common in the well-known derivation of the virial theorem [1], were not used.

From the comparison of the energy balance equation (A8) with the first law of thermodynamics (A4), it follows that the left part (A8) and the second term in its right part correspond exactly to the same terms of equation (A4). This means that it is permissible to compare both the first terms in the right-hand sides of equations (A4) and (A8), and which follows the possibility of estimating the magnitude of the entropy perturbation caused by changes in the magnitude of charges during the similarity transformation. If in (A4), when describing the change in entropy in the form $S \rightarrow \lambda_S S; \lambda_S = 1 + \delta_S, \delta_S \ll 1$ use substitution $dS \rightarrow \delta_S S$, then from the comparison with (A8) we get:

$$\delta_S = 2\varepsilon q_P;$$
$$q_P = \frac{E_P}{TS} \propto O(1) \qquad (A10)$$

At the same time, depending on the sign of the magnitudes $\varepsilon$ and $E_P$, both an increase and a decrease in entropy may possible due to the transformation of similarity associated with a change in charge. This does not contradict the law of increase, which is valid for closed systems, since the estimate of the magnitude of the entropy change obtained in (A10) is associated precisely with its change at the inner boundary of the system under consideration.

Note also that, despite the exact coincidence of the first term in the right part (A8) with the similar term in the energy balance equation given in [28], there is a difference in the justification for the appearance of this term in this form. The qualitative explanation given in [28] may proceeds from the idea that entropy remains unchanged during the considered similarity transformation. On the contrary, the justification of the representation for the first term in the right-hand side (A8) in this paper is proved on the basis of a strict quantitative assessment, which is essentially based on the assumption (A2) about the nature of the asymptotic behavior of the interaction potential, which is analogous to the asymptotic freedom in the theory of elementary particles [30]. In the argumentation [28], it is not explicitly noted that the change in the total energy in the left side of equation (A8) is due only to the presence of internal and external boundaries through which the energy flow is permissible. In this respect, it is the inner boundary, which has a zero measure of extent due to the point nature of the charge that can provide a change in entropy according to (A10). Thus, the decrease in entropy during the transformation of the similarity of the charge magnitude at the inner boundary can be interpreted as the dominance of the corresponding information factors compared in an analogy with the subject of the known studies [75]- [78].

## 2. Modification of Euler's theorem with a non-constant uniformity exponent

By the definition, a function $u_P(\vec{x})$ is a uniformity function of coordinates if the scale transformation of stretching or compression for all coordinates on which this function depends has the form

$$u_P(\lambda \hat{x}) = \lambda^n u_P(\hat{x});$$
$$\hat{x} = (\vec{x}_1; \vec{x}_2; ...; \vec{x}_N) \qquad (A11)$$

In this case, the potential energy in (A11) is the sum of all possible interactions between $N$ particles, when each of them is characterized by its position vector in a given coordinate system in which the medium is assumed to be at rest as a whole.

It is usually assumed that the value of the potential energy index $n = const$ in (A 11) is constant when Euler's theorem on uniform functions leads to a well-known representation of the virial theorem for a system of N interacting particles [1]:

$$\sum_{\alpha=1}^{N} x_i^{\alpha} \frac{\partial u_P}{\partial x_i^{\alpha}} = n u_P(\vec{x}^1, \vec{x}^2, ..., \vec{x}^N) \tag{A12}$$

On the repeated index $i = 1,2,3$ in (A 12) implies summation from 1 to 3.

At the same time, in (A11) and (A 12) there can be any constant value $\lambda$ of the scale conversion coefficient.

Consider the case when in the definition (A11) the uniformity index $n = n(\vec{x})$ depends on the coordinates. We introduce the concept of locally uniformity functions for which relation (A12) holds only for such magnitudes of the scale transformation coefficient that satisfy the additional condition $\lambda = 1 \pm \varepsilon_1; \varepsilon_1 \ll 1$. To substantiate this statement, we use the well-known method of proving Euler's theorem on homogeneous functions. To do this, we differentiate the left and right sides of relation (A11) by parameter $\lambda$. This leads to equality:

$$\sum_{\alpha=1}^{N} \frac{\partial u_P(\lambda \vec{x}_1; ...; \lambda \vec{x}_N)}{\partial (\lambda x_i^{\alpha})} \frac{\partial (\lambda x_i^{\alpha})}{\partial \lambda} = n \lambda^{n-1} u_P(\hat{x}) \tag{A13}$$

In the left part (A 13) we have $\partial(\lambda x_i^{\alpha})/\partial \lambda = x_i^{\alpha}; \partial u_P(\lambda \hat{x})/\partial(\lambda x_i^{\alpha}) = \lambda^{-1} \partial u_P(\lambda \hat{x})/\partial x_i^{\alpha}$. Therefore, from (A13) we get (since taking into account (A11) we have $\partial u_P(\lambda \hat{x})/\partial x_i^{\alpha} = \partial(\lambda^n u_P(\hat{x}))/\partial x_i^{\alpha}$) in general when $n = n(\hat{x}) \neq const$:

$$\sum_{\alpha=1}^{N} \left( x_i^{\alpha} \frac{\partial u_P(\hat{x})}{\partial x_i^{\alpha}} + u_P(\hat{x}) x_i^{\alpha} \frac{\partial n}{\partial x_i^{\alpha}} \ln \lambda \right) = n u_P(\hat{x}) \tag{A14}$$

For any $\lambda$ only in case of a constant value $n$ we obtain from (A14) the well-known statement of Euler's theorem (A12) on the uniform functions having a constant value of the exponent of uniformity in (A12). In case when this exponent of uniformity $n = n(\vec{x})$ is no longer constant, we can consider a class of locally homogeneous functions for which only in the limit $\lambda = 1 \pm \varepsilon_1; \varepsilon_1 \rightarrow 0$ of (A14) follows a modification of Euler's theorem having the following form:

$$\sum_{\alpha=1}^{N} x_{\alpha} \frac{\partial u_P(\vec{x})}{\partial x_{\alpha}} = n(\vec{x}) u_P(\vec{x}) \tag{A15}$$

In particular, when equality $n(\vec{x}) = -d(\vec{x})$ is fulfilled, the exponent of local homogeneity is related to the exponent $\tilde{d}$ of the potential of interaction of particles by the ratio (A5) in the Subsection 1 of this Appendix or equation (1) of the main text.

3. **Exact solution for the EOS.** By using the Lagrange-Sharpie method, we obtain a general implicit solution of equation (6) in the form of a transcendental algebraic equation for the function $w$:

$$a(w+1) + (a+1) \ln w = b + ax + \tilde{y};$$

$$a = \frac{\partial w/\partial x}{\partial w/\partial \tilde{y}} = \frac{w_0/w'_{0\tilde{y}} - 1}{w_0 + 1} = const; w'_{0\tilde{y}} = \left( \frac{\partial w}{\partial \tilde{y}} \right)_{x=x_0; y=y_0}$$

$$b = a(w_0+1) + (a+1) \ln w_0 - ax_0 - \tilde{y}_0 = const; w_0 = w(x=x_0; \tilde{y} = \tilde{y}_0)$$

(A16)

In order to make sure that the solution (A16) exactly satisfies equation (6), it is enough to substitute in (6) representations of two derivatives $\partial w / \partial x; \partial w / \partial \widetilde{y}$ that may be obtained from (A16) directly by differentiating (A16) with respect to the corresponding variables:

$$\frac{\partial w}{\partial x} = \frac{aw}{a(w+1)+1};$$

$$\frac{\partial w}{\partial \widetilde{y}} = \frac{w}{a(w+1)+1}$$

(A17)

In this case, the representations for the two integration constants given in (A16) are obtained by taking into account the solution given in (A16) and (A17).

In the limit $w \gg 1$ it may be interesting to compare explicit solution (8) with also exact analytical, but an implicit (similar to (A16), see in Appendix) solution of equation (7) that has the transcendent form $a(w+1)+\ln(w+1) = b_1 + ax + \widetilde{y}$, where $b_1 = const; a = const$.

Let us also consider the opposite limit $w \ll 1$ when the nonlinear Eq. (6) transforms to a linear equation and has the following form:

$$\frac{\partial w}{\partial \widetilde{y}} + \frac{\partial w}{\partial x} - w = 0$$

(A18)

Indeed, the Lagrange-Sharpie method leads to the following type of solution (A18), which is also directly obtained from solution (A16) in the limit $w \ll 1$:

$$w = w_0 \exp\left(\frac{ax + \widetilde{y}}{a+1}\right);$$

$$a = -\frac{w'_{0T}}{w'_{0T} - \frac{w_0}{T_0}} = -\frac{1}{1 - \frac{u_{P0}}{c_{V0}T_0}\ln\frac{u_{P0}}{u_0}} = const;$$

$$w_0 = w(x = 0; \widetilde{y} = 0) = \ln\frac{u_{P0}}{u_0};$$

(A19)

$$w'_{0T} = \left(\frac{\partial w}{\partial T}\right)_{x=0; \widetilde{y}=0} = \frac{c_{V0}}{u_{P0}}, c_V = \left(\frac{\partial u_P}{\partial T}\right)_V ; c_{V0} = c_V(x = 0; \widetilde{y} = 0)$$

$$d(\rho;T) = 3w_0\left(\frac{T}{T_0}\ln^{-1}\left(\frac{\rho}{\rho_0}\right)\right)^{\frac{a}{a+1}}; w_0 = \frac{\widetilde{d}_0}{3}; \widetilde{d}_0 \equiv d(x = 0; \widetilde{y} = 0)$$

The constant value $\widetilde{d}_0$ in (A19) is different from the function $d_0(T)$ that is used in the exact solution (9).

The solution (A19) of equation (A18) depends on two arbitrary constants $w_0; a$. The solution (A19) must satisfy condition $w \ll 1$, which in case $a = -|a|; 0 < |a| < 1$ has the form

$$\frac{\widetilde{d}_0}{3}\left(\frac{T_0}{T}\ln\frac{\rho}{\rho_0}\right)^{\frac{|a|}{1-|a|}} \ll 1$$ . For example, for fixed values $\widetilde{d}_0$ and $T / T_0$ the specified condition is fulfilled in the limit of small compression values $\delta = \rho / \rho_0$, when $\delta \rightarrow 1$.

On the other hand, the general solution of equation (A18) can be obtained up to an arbitrary function, since it is a partial differential equation of the first order (in the limit $w \ll 1$ for the case of small compressions at a finite functions $d(\rho;T)$ and $F_0 \propto O(1)$ ):

$$w = e^{\tilde{y}} \tilde{F}_0(x - \tilde{y});$$

$$w \equiv \frac{d(\rho;T)}{3} \ln \frac{\rho}{\rho_0}; w \ll 1, \rho - \rho_0 \ll \rho_0; \qquad (A20)$$

$$d(\rho;T) = \tilde{F}_0\left( \ln\left(\frac{T}{T_0}\right) - \ln\ln\left(\frac{\rho}{\rho_0}\right)\right)$$

In Eq. (A20) $F_0$-is an arbitrary smooth function and only in case $\tilde{F}_0(x - \tilde{y}) = w_0 \exp\left(\frac{a(x - \tilde{y})}{a + 1}\right)$ the solution (A19) is a special case of the general solution (A20).

Let us use the EOS (3) in the simple case of exact solution (A19), to estimate a value $B = \lim_{\rho \to 0}\left(\frac{\partial(pm/kT\rho)}{\partial \rho}\right)_T$ [5], [6] that has the following form:

$$B = \frac{u_0 m \tilde{d}_0^2}{9kT} \lim_{\rho \to 0} \frac{\exp(F + F_1)}{(1+a)\rho \ln^2 \rho/\rho_0}\left[-\frac{3a}{\tilde{d}_0} + \exp F\right];$$

$$F = \frac{ax + \tilde{y}}{a + 1}; F_1 = \frac{\tilde{d}_0}{3} \exp F; x = \ln\frac{T}{T_0}; \tilde{y} = \ln\ln\frac{\rho}{\rho_0}; \qquad (A21)$$

$$a = -\frac{1}{1 - \frac{\tilde{d}_0}{(T\partial(d(x;\tilde{y})\ln(\rho/\rho_0)/dT)_0}}; \tilde{d}_0 \equiv d(x = 0; \tilde{y} = 0)$$

In (A21), there is a possible case with $\tilde{d}_0 < 0$ when $\delta = \rho/\rho_0 = \eta < 1$ as in the solution (35). In Eq. (A21), the finite value of second virial coefficient in the limit $\rho \to 0$ can be obtained only in case when $\lim_{\rho \to 0}\left(\delta^{H(x;a)}/\ln^{S_\beta} \delta\right) < \infty; \beta = 1;2; H = \frac{\tilde{d}_0 x^{\frac{a}{a+1}}}{3(a+1)} - 1; S_1 = \frac{2a+1}{a+1}; S_2 = \frac{2a}{a+1}$.

Thus, it follows $B = 0$ from (A21) when for the positive values of the parameter $a/\tilde{d}_0 > 0$ the square bracket in (A21) is zeroed. This gives the estimate for the Boyle temperature in the simple analytical form:

$$T_B/T_0 = (3a/\tilde{d}_0)^{\frac{a+1}{a}} |\ln \delta|^{-1/a}; a/\tilde{d}_0 > 0 \qquad (A22)$$

It follows that $B > 0$ in (A21) at a temperature of $T > T_B$ for $T_B$ from (A22) in case $a > -1$. The Boyle temperature in (A22) has a weak logarithmic dependence on the density for any finite density values. Similarly, from (A21) for the temperature $T = T^\pm$ at which the first derivative vanishes $(\partial B/\partial T)_\rho = 0$, we obtain the following representation:

$$\frac{T^{\pm}}{T_0} = \left( \ln \frac{\rho}{\rho_0} \right)^{-1/a} \left[ \frac{a^2 + 1 - a \pm \sqrt{\left( a^2 + 1 - a \right)^2 - 4a^2}}{2a\tilde{d}_0 / 3} \right]^{\frac{a+1}{a}}; T^+ = T_{\min}; T^- = T_{\max}$$

$$a \geq a_+ = 1 + G_+ \approx 2.618033 \ or : a \leq a_- = 1 - G_- \approx 0.381966, \qquad \text{(A23)}$$

$$G_{\pm} = \frac{\sqrt{5} \pm 1}{2}$$

In (A 23) $G_{\pm}$ - are the well-known numbers associated with the golden ratio and with a sequence of Fibonacci's numbers [79]-[82].

The inequality $(\partial^2 B / \partial T^2)_{\rho} < 0$ and, accordingly, the maximum of the function $B$ at temperature $T = T_{\max} = T^-$ (A23) take place when in (A23) one takes sign minis. In the opposite case, the minimum value of function $B$ takes place at the temperature $T = T_{\min} = T^+$.

In (A23) there are represented the restrictions on the parameter $a$, under which the root expression in (A23) is not negative. In cases $a_- > a > 0$, $a < -1$ and $a > a_+$, relation $T_{\min} > T_{\max}$ is obtained from (A23). And only in case $-1 < a < 0$, relation $T_{\max} > T_{\min}$ is valid.

The well-known condition $T_{\max} > T_B$ [5], [6] is also valid only in any of the following cases: $a > 1 + G_+$, $-1 < a < 1 - G_-$, or $a < -G_+$. The possibility of fulfilling the opposite inequality $T_{\max} < T_B$ according to (A23) and (A22) is permissible only under condition $-G_+ < a < -1$. Note also that the inequality $T_B < T_{\min}$ holds if one of conditions $a < -1$ or $0 < a < 1 - G_-$ is satisfied. The opposite inequality $T_B < T_{\min}$ is realized if one of conditions $-1 < a < 0$ or $a > 1 + G_+$ is met.

It is possible in future to make compare these estimates on the basis of an exact analytical solution for the corresponding quantities (A22) and (A23), assuming that equality $T / T_0 = T^*$ in (A22) and (A23) is fulfilled. In particular, as in [62], from (A22) and (A23) follows the fulfillment of the necessary ratio $T_{\max}^* > T_B^*$. The last relation can take place only if $-1 < a < 0$ when only one branch of the solution $T_{\max} = T^-$ is selected in (A23).

### Appendix B The closure problem in the compressible hydrodynamics

Let us consider equations of hydrodynamics of a viscous compressible medium, which have the following form:

$$\frac{\partial u_i}{\partial t} + u_j \frac{\partial u_i}{\partial x_j} = -\frac{1}{\rho} \frac{\partial}{\partial x_i} \left( p - \left( \varsigma + \frac{\eta}{3} \right) div \vec{u} \right) + \frac{\eta}{\rho} \Delta u_i \qquad \text{(B1)}$$

$$\frac{\partial \rho}{\partial t} + \frac{\partial (\rho u_i)}{\partial x_i} = 0 \qquad \text{(B2)}$$

Based on equations (B1) and (B2), an expression can be obtained for the time derivative of the integral kinetic energy of the flow of a viscous compressible medium. According to the definition, this energy has the form [42]:

$$E_K = \frac{1}{2} \int d^3 x \rho \vec{u}^2 \qquad \text{(B3)}$$

Differentiating the energy value in (B3) by time when directly using (B1) and (B2), we obtain the following equation of the kinetic energy balance of the flow of a viscous compressible medium:

$$\frac{dE_K}{dt} = -\eta \int d^3x \left(\frac{\partial u_i}{\partial x_j}\right)^2 + \int d^3x \, div\vec{u} \left( p - \left(\varsigma + \frac{\eta}{3}\right) div\vec{u} \right) \qquad (B4)$$

The balance equation (B4) (in the limit of zero shear $\eta$ and volumetric viscosities $\varsigma$) coincides with the one given in G. Lamb's book [83] for the case of compressible flows. It is equation (B4) that gives a generalization for the energy balance equation of the incompressible medium flow, given in [42] in the form of equation (16.3) when compressibility must be taken into account.

At the same time, in [42] it is stated that, taking into account the compressibility of the medium, the generalization of equation (16.3) is equation (79.1), which for the isothermal case is expressed as follows:

$$\frac{dE_K}{dt} = -\varsigma \int d^3x \, div^2\vec{u} - \frac{\eta}{2} \int d^3x \left(\frac{\partial u_i}{\partial x_j} + \frac{\partial u_j}{\partial x_i} - \frac{2}{3}\delta_{ij} div\vec{u}\right)^2 \qquad (B5)$$

At the same time, the kinetic energy balance equation of the compressible medium given in [42] in the form of (B5), in contrast to (B4), no longer coincides with the balance equation given in G. Lamb's book in the zero-viscosity limit. As a result, equation (B5) differs from the balance equation (B4) obtained directly from the hydrodynamic equations (B1) and (B2). This means that equation (B5) contradicts equations (B1), (B2) and cannot be a generalization of the integral kinetic energy balance equation for the case of a compressible medium.

We show that the indicated contradiction of equation (B5) to equations of the hydrodynamics of a compressible medium (B1), (B2) is the use of the thermodynamic relation (see below (B9)) for the pressure gradient used in the derivation (B5) based on the integral entropy balance equation. Indeed, in order to derive equation (B5) in [42], the integral entropy balance equation is used, which is obtained from the condition of conservation of the total integral energy, representing the sum of the integral kinetic and integral internal energy in the following form:

$$E = E_K + E_I;$$
$$E_I = \int d^3x \rho \varepsilon_I \qquad (B6)$$

For this purpose, in [42], based on equations of hydrodynamics of a compressible medium (B1) and (B2), the rate of change in time for the total energy density is obtained, which has the following form:

$$\frac{\partial}{\partial t}\left(\frac{\rho u^2}{2} + \rho\varepsilon_I\right) = -\left(\frac{u^2}{2} + \frac{p}{\rho} + \varepsilon_I\right) div(\rho\vec{u}) - \rho(\vec{u}\vec{\nabla})\frac{u^2}{2} - \vec{u}\vec{\nabla}p + \rho T \frac{\partial s}{\partial t} + u_i \frac{\partial \sigma'_{ik}}{\partial x_k};$$

$$\sigma'_{ik} = \eta\left(\frac{\partial u_i}{\partial x_k} + \frac{\partial u_k}{\partial x_i} - \frac{2}{3}\delta_{ik} div\vec{u}\right) + \varsigma\delta_{ik} div\vec{u} \qquad (B7)$$

For the derivation (B7) in [42], an equation derived from the first law of thermodynamics $d\varepsilon_I = Tds + pd\rho/\rho^2$ and the continuity equation (B2) was used, which has the following form:

$$\frac{\partial \varepsilon_I}{\partial t} = T \frac{\partial s}{\partial t} - \frac{p}{\rho^2} div(\rho\vec{u}) \qquad (B8)$$

In addition (and this is most significant in terms of understanding the difference between (B5) and (B4)), in [42], a thermodynamic relation $d(\varepsilon_I + \frac{p}{\rho}) = Tds + dp/\rho$ was used to transform into (B7) term $\vec{u}\vec{\nabla}p$, containing a pressure gradient, leading to a representation for the pressure gradient in the following form:

$$\vec{\nabla}p = \rho\vec{\nabla}(\varepsilon_I + \frac{p}{\rho}) - \rho T\vec{\nabla}s \qquad (B9)$$

After substituting (B9) into (B7) and from the condition of conservation of the total integral energy $dE/dt = 0$ in [42], the integral entropy $S = \int d^3x \rho s$ balance equation is obtained (see (49.6) in [42]), which in the isothermal case and for constant coefficients of shear and bulk (second) viscosity has the following form:

$$T\frac{dS}{dt} = \frac{\eta}{2}\int d^3x \left(\frac{\partial u_i}{\partial x_k} + \frac{\partial u_k}{\partial x_i} - \frac{2}{3}\delta_{ik}div\vec{u}\right)^2 + \varsigma\int d^3x div^2\vec{u} \qquad (B10)$$

Further, in [42] from (B10), in order to obtain the kinetic energy balance equation in the form (B5), a well-known relation was used, expressed for any mechanical system in the isothermal case $T = T_0 = const$, which has the following form:

$$T_0\frac{dS}{dt} = -\frac{dE_K}{dt} \qquad (B11)$$

Let us show that relation (B11) can be fulfilled with the independent derivation of the balance equation (B4) for the integral kinetic energy and for the balance equation of the integral entropy obtained as in [42], based on the condition of conservation of the integral total energy, but without using the thermodynamic relation (B9) for the pressure gradient. To do this, instead of (B8) we will further use the following equation:

$$\frac{\partial\rho\varepsilon_I}{\partial t} = T\frac{\partial(\rho s)}{\partial t} + \Phi\frac{\partial\rho}{\partial t} \qquad (B12)$$

In this case, the Gibbs thermodynamic potential is used, which has the following form:

$$\Phi = \mu_{ch}N = \varepsilon_I + \frac{p}{\rho} - Ts;$$

$$d\varepsilon_I = Tds + \frac{p}{\rho^2}d\rho; \qquad (B13)$$

$$d\Phi = \frac{dp}{\rho} - sdT$$

As a result, we obtain the integral entropy balance equation from the condition of conservation of the total integral energy [33], [71]:

$$\frac{dE}{dt} = -\int d^3x div\vec{J}_E + \frac{dS}{dt} - B = 0; S = \int d^3x \rho s; \qquad (B14)$$

$$\vec{J}_E = \vec{u}\left(\rho\left(\frac{\vec{u}^2}{2} + \Phi + Ts\right) + p - \left(\varsigma + \frac{\eta}{3}\right)div\vec{u}\right) - \frac{\eta}{2}\vec{\nabla}\vec{u}^2$$

$$\frac{dS}{dt} = B;$$

$$T_0 B = \eta \int d^3 x \left( \frac{\partial u_i}{\partial x_j} \right)^2 - \int d^3 x \, div \vec{u} \left( p - \left( \varsigma + \frac{\eta}{3} \right) div \vec{u} \right) - \int d^3 x \Phi \, div(\rho \vec{u}) \qquad (B15)$$

After comparing (B15) and (B4), we obtain a generalization of relation (B11) in the following form:

$$\frac{dS}{dt} = -\frac{1}{T_0} \frac{dE_K}{dt} - \frac{1}{T_0} \int d^3 x \Phi \, div(\rho \vec{u}) \qquad (B16)$$

Representation (B16) under condition $\Phi = \mu_{ch} N = \Phi_0 = const$ matches (B11). In this form, it is possible to see a generalization as a case of a variable number of particles $N$ and a variable chemical potential $\mu_{ch}$ when an additional condition in is required to fulfill relation (B11):

$$\int d^3 x \Phi \, div \vec{u} = 0 \qquad (B17)$$

Thus, the ratio (B11) is obtained based on equations of hydrodynamics (B1), (B2) only when using the thermodynamic ratio (B12) without the need to apply the representation (B9) for the pressure gradient. Therefore, due to the direct contradiction of equation (B5) and equations of hydrodynamics of a compressible medium (B1), (B2), which is evident, it is possible to make a statement about a similar contradiction of the integral entropy balance equation (B10), if a fundamental relation (B11) is valid [42]. In turn, this necessarily leads to the conclusion that it is unacceptable to use the thermodynamic relation (B9) to describe the dynamic pressure gradient included in the hydrodynamic equations (B1) when deriving the integral entropy balance equation.

In this regard, it is necessary to revise the applicability of the representation for pressure following from the thermodynamic EOS in Eq. (B1) for the closure the hydrodynamics equations in all cases where this could be avoided, as in the above derivation of the integral entropy balance equation (B15). In contrast to Eq. (B10) (see also (49.6) in [42]), the EOS (B15) does not contradict equations of hydrodynamics (B1), (B2) in their combination with equation (B11), that is suggested to be valid for any mechanical system [42].

### Appendix C. The Grüneisen parameter

1. Using the definition of the Grüneisen parameter as $\frac{1}{\rho} \left( \frac{\partial p}{\partial u} \right)_{\rho=const} = \frac{1}{\rho} \left( \frac{\partial p}{\partial T} \right)_{\rho=const} \left( \frac{\partial u}{\partial T} \right)_{\rho=const}^{-1} = \Gamma$ from (27) and (28), we obtain this parameter in the Hugoniot shock adiabat as follows

$$\Gamma = \frac{2}{\delta - 1}, |d| \neq 2$$
$$\Gamma = \frac{2}{3}, |d| = 2 . \qquad (C1)$$

The Grüneisen coefficient defined in (26) coincides with (C1) only when the condition imposed on the compression in the form $\delta \to \delta_d \equiv 1 + 6/|d|$ when (31) is fulfilled. The dependence of the Grüneisen parameter on compression in (C1) qualitatively coincides with the estimate of this parameter presented in [45], which monotonically decreases with increasing compression at sufficiently large compression values (see Fig. 12(b) in [45] and Fig. 11 in [46]).

To obtain representation (32) for the Grüneisen parameter in the form $\Gamma = c_T^2 \beta / c_V$, we use the following form of the specific heat capacity $c_V$, which was obtained from the virial theorem in the form (15), and relation $\alpha \frac{p}{\rho} = \beta c_T^2$ [see (C8) below]

$$c_V = \left(\frac{\partial u}{\partial T}\right)_{\rho=const} = \alpha \frac{p}{\rho}\frac{3}{|d|} + 3\frac{k}{m_{H_2O}}\left(\frac{1}{2} - \frac{1}{|d|}\right); \alpha = \frac{1}{p}\left(\frac{\partial p}{\partial T}\right)_{\rho=const} \qquad (C2)$$

$$c_p = \left(\frac{\partial(u+p/\rho)}{\partial T}\right)_{p=const} = \beta\frac{p}{\rho}\left(1 + \frac{3}{|d|}\right) + 3\frac{k}{m_{H_2O}}\left(\frac{1}{2} - \frac{1}{|d|}\right); \ \beta = -\frac{1}{\rho}\left(\frac{\partial\rho}{\partial T}\right)_{p=const} . \quad (C3)$$

Also, from (C2) and (C3) the adiabatic exponent reads

$$\gamma = \frac{c_p}{c_V} = \frac{a_1 + g}{b_1 + g}, \quad a_1 = \beta\frac{p}{\rho}\left(1 + \frac{3}{|d|}\right), \ b_1 = \alpha\frac{3p}{|d|\rho}; g = \frac{3k}{m_{H_2O}}\left(\frac{1}{2} - \frac{1}{|d|}\right). \quad (C4)$$

Using (C4) and the condition on the adiabatic index $\gamma > 1$, we obtain the following limitation on parameter $d$:

$$|d| > 3\left(\frac{\alpha}{\beta} - 1\right) = 3\left(\frac{\rho c_T^2}{p} - 1\right) . \qquad (C5)$$

Relation $\alpha \frac{p}{\rho} = \beta c_T^2$ is obtained by considering the known relations [1] (see Formulas (16.9) and (16.10) in [1]):

$$c_p - c_V = -T(\partial V/\partial T)_{p=const}^2 / (\partial V/\partial p)_{T=const} \qquad (C6)$$
$$c_p - c_V = -T(\partial p/\partial T)_{V=const}^2 / (\partial p/\partial V)_{T=const}. \qquad (C7)$$

After equating the right-hand sides of (C6) and (C7),

$$\frac{p^2\alpha^2}{\rho^2\beta^2} = \left(\frac{\partial p}{\partial\rho}\right)_{T=const}^2 = c_T^4 \qquad (C8)$$

From (C8), we obtain relation $\alpha \frac{p}{\rho} = \beta c_T^2$. Relation (C8) includes the value of the isothermal velocity of sound $c_T$ [42]. This value for the isothermal EOS of water (19) has the form

$$c_T^2 = c_{0T}^2 \delta^{n-1} . \qquad (C9)$$

As a result, using (C9) and considering (C4), we obtain for $\gamma = c_p/c_V > 1$

$$|d| > 3\left(\frac{\rho_0 c_0^2 \delta^n}{p} - 1\right) \qquad (C10)$$

In case where the pressure in (C10) satisfies the isothermal EOS (19), from (C10) and (19) in the limit $\delta^n \gg 1; p \gg p_0$, relation $p \cong \rho_0 c_0^2 \delta^n / n$ becomes valid and condition (C10) is reduced to the following form

$$n < 1 + \frac{|d|}{3}; o\,r\,|d| > 3(n-1) . $$

$$(C11)$$

## Data availability

The data supporting the findings of this study are available from the corresponding author upon reasonable request.